\documentclass[final,5p,times,sort&compress]{elsarticle}

\usepackage{graphicx}
\usepackage{amsmath,amssymb,amsfonts}
\usepackage{bm}

\usepackage{multirow}

\journal{Annals of Physics}

\begin{document}

\begin{frontmatter}

\title{Quantum interference within the complex quantum Hamilton-Jacobi
formalism}

\author[UT]{Chia-Chun~Chou}
 \ead{chiachun@mail.utexas.edu}
\author[ESP]{\'Angel~S.~Sanz}
 \ead{asanz@imaff.cfmac.csic.es}
\author[ESP]{Salvador~Miret-Art\'es}
 \ead{s.miret@imaff.cfmac.csic.es}
\author[UT]{Robert~E.~Wyatt}
 \ead{wyattre@mail.utexas.edu}
\address[UT]{Institute for Theoretical Chemistry and Department of
Chemistry and Biochemistry, The University of Texas at Austin,
Austin, Texas 78712, USA}
\address[ESP]{Instituto de F\'{\i}sica Fundamental, Consejo Superior
de Investigaciones Cient\'{\i}ficas, Serrano 123, 28006 Madrid,
Spain}

\begin{abstract}
Quantum interference is investigated within the complex quantum
Hamilton-Jacobi formalism. As shown in a previous work [Phys.\ Rev.\
Lett.\ {\bf 102}, 250401 (2009)], complex quantum trajectories
display helical wrapping around stagnation tubes and hyperbolic
deflection near vortical tubes, these structures being prominent
features of quantum caves in space-time Argand plots. Here, we
further analyze the divergence and vorticity of the quantum momentum
function along streamlines near poles, showing the intricacy of the
complex dynamics. Nevertheless, despite this behavior, we show that
the appearance of the well-known interference features (on the real
axis) can be easily understood in terms of the rotation of the nodal
line in the complex plane. This offers a unified description of
interference as well as an elegant and practical method to compute
the lifetime for interference features, defined in terms of the
average wrapping time, i.e., considering such features as a resonant
process.
\end{abstract}


\end{frontmatter}


\section{\label{sec:introduction} Introduction}

Quantum interference is an observable effect arising from the coherent
superposition of quantum probability amplitudes.
It is involved in a very wide range of important applications, such as
superconducting quantum interference devices or SQUIDs \cite{scalapino,%
friedman}, coherent control of chemical reactions \cite{paul},
atomic and molecular interferometry
\cite{berman,itano,urena1,urena2}, and Talbot/Talbot-Lau
interferometry with relatively heavy particles (e.g., Na atoms
\cite{chapman2} and Bose-Einstein condensates \cite{deng}). Indeed,
possibly one of the main practical applications of interference
nowadays is in Bose-Einstein condensate (BEC) interferometry \cite{yoo,%
dalibard,pethick} due to its potential use in applications, such as
sensing, metrology or quantum information processing.
Thus, since the first experimental evidence of interference between
two freely expanding BECs was observed \cite{ketterle}, an increasing
amount of work, both theoretical and experimental, can be found in the
literature \cite{pritchard,zhang:070403,cederbaum:110405,chip0,chip1,%
chip2,chip3}.  Basically, in this type of interferometry there are
three steps. First, the atomic cloud is cooled in a magnetic trap
until condensation takes place; second, the BEC is split coherently
by means of radio-frequency \cite{chip1} or microwave fields
\cite{chip2}; and third, the double-well-like trapping potential is
switched off and the two parts of the BEC are allowed to interfere
by free expansion (see, for example, Fig.~6 in Ref.~\cite{chip3} as
an illustration of a real experimental outcome). Apart from the
practical applications mentioned before, interference plays also an
important role when dealing with multipartite entangled systems as
an indicator of the loss of coherence induced by the interaction
between the different subsystems, commonly referred as
``Schr\"odinger cat states'' \cite{schro}. However, very little
attention beyond the implications of the superposition principle has
been devoted to understanding quantum interference at a more
fundamental level.

Bohmian mechanics provides an alternative interpretation to quantum
mechanics \cite{bohm1,bohm2,HollandBook}. As an \emph{analytical}
approach, this formulation has been used to analyze atom-surface
scattering \cite{ASSanz,ASSanz2,ASSanz4}, the quantum Talbot effect
\cite{ASSanz6}, quantum nonlocality \cite{ASSanz7} or quantum
interference \cite{ASSanz9}. As a \emph{synthetic} approach, the
quantum trajectory method (QTM) \cite{LopreoreWyatt1} has been
developed as a computational implementation to the hydrodynamic
formulation of quantum mechanics to generate the wave function by
evolving ensembles of quantum trajectories \cite{WyattBook}.
However, computational difficulties resulting from interference
effects are encountered in regions where the wave packet is
reflected from the barrier and nodes or quasi-nodes occur.  Thus, a
bipolar counter-propagating decomposition approach for the total
wave function has been developed to overcome numerical instabilities
due to interferences
\cite{Poirier1,Poirier2,Poirier3,Poirier5,Poirier6,Poirier7,Poirier8}.

Here, a more detailed analysis of quantum interference within the
complex quantum Hamilton-Jacobi formalism \cite{LP1,LP2} is presented.
The complex quantum trajectory method has been applied to stationary
problems \cite{MJohn1,MJohn2,CDYang1,CDYang2,CDYang3,Chou3a,Chou3b}
and to one-dimensional and multi-dimensional wave packet scattering
problems \cite{Tannor2a,Tannor2b,Tannor2c,Tannor3a,Tannor3b,Rowland1a,%
Rowland1b,Rowland1c,Rowland2a,Rowland2b,Rowland2c}. The purpose of
this work is to investigate how information can be extracted from
this complex formulation regarding interference. Recently, quantum
interference associated with superpositions of Gaussian wave packets
was thoroughly analyzed using both the standard quantum-mechanical
and Bohmian approaches in real space \cite{ASSanz8, ASSanz9}. In
spite of its simplicity, such superpositions are experimentally
realizable in atom interferometry
\cite{haensel-a,haensel-b,haensel-c,haensel-d,haensel-e}, for
example.  In contrast to conventional quantum mechanics, Bohmian
mechanics offers a trajectory-based understanding of quantum
interference. The interference of two wave packets in one real
coordinate leads to the formation of nodal structure, and the
quantum potential near these nodes forces these trajectories to
avoid these regions and to exhibit laminar flow in space-time plots
\cite{ASSanz8}. In contrast, within the complex quantum
Hamilton-Jacobi formalism, the collinear collision of two Gaussian
wave packets demonstrated caustics and vortical dynamics in the
complex plane \cite{ASSanz8,Chou9}.  This complicated trajectory
dynamics cannot be found in Bohmian mechanics unless two or more
real coordinates are introduced.

When systematically analyzed for local structures of the quantum
momentum function (QMF) and the P\'{o}lya vector field (PVF) around
its characteristic points \cite{Chou6,Chou7}, complex quantum
trajectories display \emph{helical wrapping} around stagnation tubes
and \emph{hyperbolic deflection} around vortical tubes \cite{Chou9}.
This intriguing topological structure is formed by these tubes and
gives rise to \emph{quantum caves} in space-time Argand plots.
Quantum interference thus leads to the formation of quantum caves
and the appearance of the topological structure mentioned before. In
contrast with quantum trajectories, P\'{o}lya trajectories display
\emph{hyperbolic deflection} around stagnation tubes and
\emph{helical wrapping} around vortical tubes. The QMF divergence
and vorticity characterize the turbulent flow of trajectories,
determining the so-called {\it wrapping time} \cite{Chou9} for an
individual trajectory. Moreover, it is shown that both the PVF
divergence and vorticity vanish except at poles, thus the PVF
describing an incompressible and irrotational flow.

Trajectories launched from different positions wrap around the same
stagnation tubes. Hence, the circulation of trajectories can be
viewed as a resonance process, from which one can obtain a natural
way to define a ``lifetime'' for the interference. This information
could be therefore used with practical purposes to analyze, explain
and understand experiments where interference is the main physical
process or mechanism, as in those described above in this Section.
The rotational dynamics of the nodal line arising from the
interference of wave packets in the complex plane thus offers an
elegant method to compute the lifetime for the interference features
observed on the real axis.

We consider two cases of head-on collision of two Gaussian wave
packets, which depend on the relative magnitude between the
propagation velocity and the spreading rate of the wave packets
\cite{ASSanz9}. In the case where the relative propagation velocity
is larger than the spreading rate of the wave packets, an average
wrapping time is calculated to provide a lifetime for the
interference features, while the rotational dynamics of the nodal
line in the complex plane explains the transient appearance of the
interference features observed on the real axis. In the case where
the relative propagation velocity is approximately equal to or
smaller than the spreading rate of the wave packets, the rapid
spreading of the wave packets leads to a distortion of quantum
caves.  The infinite survival of interference features in this case
implies that the wrapping time becomes infinity. However, the
rotational dynamics of the nodal line clearly explains the
persistent interference features observed on the real axis. In both
cases, the interference features are observed on the real axis when
the nodal line is near the real axis; therefore, in contrast to
conventional quantum mechanics, the rotational dynamics of the nodal
line in the complex plane provides a fundamental interpretation of
quantum interference.

The organization of this work is as follows. To be self-contained,
in Sec.~\ref{sec:theory} we briefly describe the complex quantum
Hamilton-Jacobi formalism as well as the QMF local structures and its
associated PVF near characteristic points. In Sec.~\ref{sec:QI}, first
we present theoretical analysis of the Gaussian wave-packet
head-on collision on the real axis and in the complex plane, and
then demonstrate the interpretation of quantum interference with two
cases in the framework of the complex quantum trajectory method.
Finally, we present a summary and discussion in
Sec.~\ref{sec:Conclusions}.


\section{\label{sec:theory} Theoretical formulation}


\subsection{\label{sec:CQHJF}Complex quantum Hamilton-Jacobi
formalism}

Substituting the complex-valued wave function expressed by
$\Psi(x,t)=\exp\left[iS(x,t)/\hbar\right]$ into the time-dependent
Schr\"{o}dinger equation, we obtain the quantum Hamilton-Jacobi
equation (QHJE) in the complex quantum Hamilton-Jacobi formalism,
\begin{equation} -\frac{\partial S}{\partial
t}=\frac{1}{2m}\left(\frac{\partial S}{\partial
x}\right)^2+V(x)+\frac{\hbar}{2mi}\frac{\partial^2 S}{\partial x^2},
\label{TDQHJE}
\end{equation}
where $S(x,t)$ is the complex action and the last term is the
complex quantum potential.
In real space, the QMF is defined by $ p(x,t)=\partial S(x,t)/
\partial x$, which, within the complex quantum Hamilton-Jacobi
formalism, is analytically continued to the complex plane by extending
the real variable $x$ to a complex variable $z=x+iy$.
The same is done with other relevant functions, such as the wave
function and the potential energy.
Quantum trajectories in complex space are then determined by solving
the guidance equation
\begin{equation}
\frac{dz}{dt}=\frac{p(z,t)}{m} ,
\label{TDtraj1}
\end{equation}
where time remains real-valued and the (complex) QMF is expressed in
terms of the (complex) wave function (through the complex action) as
\begin{equation}
p(z,t)=\frac{\hbar}{i}\frac{1}{\Psi(z,t)}
\frac{\partial\Psi(z,t)}{\partial z}. \label{nonstationaryQMF}
\end{equation}
From this equation, nodes in the wave function correspond to QMF
{\it poles}. Moreover, those points where the first derivative of
the wave function vanishes will correspond to QMF {\it stagnation
points}.


\subsection{\label{sec:LSQMF} Local structures of the quantum momentum
function and its P\'{o}lya vector field}

The dynamics of complex quantum trajectories is guided by the wave
function through the QMF.  The complex-valued QMF can be regarded as
a vector field in the complex plane.  The trajectory dynamics is
significantly influenced by both the QMF stagnation points and poles.
Streamlines near a QMF stagnation point may spiral into or
out of it, or they may become circles or straight lines.  The QMF
near a pole displays East-West and North-South hyperbolic structure
\cite{Chou6,Chou7}.  These local structures around these
characteristic points have also been observed for the
one-dimensional stationary scattering problems including the Eckart
and the hyperbolic tangent barriers \cite{Chou3a,Chou3b}.

The PVF of a complex vector field $f(z)$ is defined by its complex
conjugate $f^*(z)$ \cite{Polya,Braden,Needham}.
Thus, the PVF associated with the QMF, $p(z,t)=p_x+ip_y$, is given by
the vector field ${\bf P}(z,t)=p_x-ip_y = (p_x,-p_y)$.
This new vector field provides a simple geometrical and physical
interpretation for complex circulation integrals
\begin{equation}
\oint_C p(z) dz = \oint_C {\bf P} \cdot d \ell
  + i \oint_C {\bf P} \cdot d {\bf n} ,
 \label{CIQMF}
\end{equation}
where $C$ denotes a simple closed curve in the complex plane, $d
\ell = (dx,dy)$ is tangent to $C$ and $d {\bf n} = (dy,-dx)$ is
normal to $C$ and pointing outwards. The real part of the integral
in Eq.~(\ref{CIQMF}) gives the total amount of work done in moving a
particle along a closed contour $C$ subjected to ${\bf P}$, while
its imaginary part gives the total flux of the vector field across
the closed contour \cite{Needham}.

In Bohmian mechanics, quantum vortices form around nodes in two or more
real coordinates \cite{Dirac,McCullough1,McCullough2,McCullough3,%
Hirschfelder1,Hirschfelder2,Hirschfelder3,Hirschfelder4,Hirschfelder5,%
ASSanz2,ASSanz3}.
Analogously, in the complex plane quantum vortices form around nodes of
the wave function, the quantized circulation integral arising from the
discontinuity in the real part of the complex action \cite{Chou6,Chou7}.
The PVF of a complex function contains exactly the same information as
the complex function itself, but it is introduced to interpret the
circulation integral in terms of the work and flux of its PVF along
the contour.
Moreover, the PVF is the tangent vector field of contours for the
complex-extended Born probability density \cite{Chou8a,Chou8b}.
Streamlines near a PVF stagnation point display rectangular hyperbolic
structure, while streamlines near a PVF pole become circles enclosing
the pole \cite{Chou6,Chou7}.
Local structures or streamlines for the QMF and its associated PVF are
summarized in Table~\ref{tab:table1}.

\begin{table}
\caption{\label{tab:table1} Local structures or streamlines for the
QMF and the associated PVF near a stagnation point or a pole.}
 \begin{center}
 \begin{tabular}{c|c|c}
  \hline \hline
   & Stagnation point & Pole \\ \hline
  \multirow{2}{*}{QMF} & Spirals, circles, & East-West and
   North-South  \\
   & or straight lines & opening hyperbolic flow \\ \hline
  \multirow{2}{*}{PVF} & Rectangular & \multirow{2}{*}{Circular flow} \\
   & hyperbolic flow &\\ \hline \hline
 \end{tabular}
 \end{center}
\end{table}


\subsection{\label{sec:} Approximate quantum trajectories around a
stagnation point}

Since the QMF is generally time-dependent, we can determine
approximate complex quantum trajectories around a stagnation point
$(z_0, t_0)$ in spacetime.  We expand the QMF in a Taylor series
around a stagnation point
\begin{equation}
p(z,t)=\left(\frac{\partial p}{\partial
z}\right)_0(z-z_0)+\left(\frac{\partial p}{\partial
t}\right)_0(t-t_0)+\cdots, \label{ThreeDQMFTS}
\end{equation}
where $p(z_0,t_0)=0$ has been used and the partial derivatives are
evaluated at the stagnation point.  Substituting this equation into
Eq.~(\ref{TDtraj1}), we obtain a first-order nonautonomous complex
ordinary differential equation for approximate quantum trajectories
\begin{equation}
\frac{dz}{dt}=\frac{1}{m}\left(\alpha z+\beta t\right),
\end{equation}
where $(\partial p/\partial z)_0=\alpha$ and $(\partial p/\partial
t)_0=\beta$ have been used and the origin has been moved to the
stagnation point.  General solutions for the nonautonomous linear
differential equation are given by \cite{Hirsch}
\begin{eqnarray}
 z(t)& = & e^{\alpha t/m} \left(z(0) + \frac{\beta}{m}
  \int_0^t e^{- \alpha s/m} sds\right) \nonumber \\
     & = & z(0) e^{\alpha t/m} + \frac{\beta m}{\alpha^2}
  \left[e^{\alpha t/m} - \left(1 + \frac{\alpha}{m} \ \! t
   \right) \right] ,
 \label{ThreeDfirstorderTJ}
\end{eqnarray}
where $z(0)$ is the starting point of a local approximate quantum
trajectory.  If we consider complex quantum streamlines at a
specific time $t_s$, the QMF $p(z,t_s)$ does not depend on time.
Hence, the partial derivative of the QMF with respect to time is
equal to zero, $(\partial p/\partial t)_0=\beta=0$.  Thus, the
general solution in Eq.~(\ref{ThreeDfirstorderTJ}) gives the
approximate quantum streamlines near the stagnation point.


\subsection{\label{sec:DivVort} Divergence and vorticity of the quantum
momentum field and its P\'{o}lya vector field}

The QMF first derivative contains the information about the
divergence and vorticity of the quantum fluid in the complex plane
\cite{Chou9}.
This can be shown as follows.
The QMF {\it divergence}, which describes the local
expansion/contraction of the quantum fluid, is given by
\begin{equation}
 \Gamma =  \nabla \cdot p = \frac{\partial p_x}{\partial x}
  + \frac{\partial p_y}{\partial y}.
\end{equation}
Analogously, the \emph{vorticity} describing the local rotation of the
quantum fluid is defined by the QMF curl,
\begin{equation}
 \Omega = \left| \nabla \times p \right| =
   \frac{\partial p_y}{\partial x} - \frac{\partial p_x}{\partial y} .
\end{equation}
Since the QMF is analytically extended to the complex plane, we use
the Cauchy-Riemann equations to write the QMF first derivative as
\begin{equation}
\frac{\partial p}{\partial z}=\frac{\partial p_x}{\partial
x}+i\frac{\partial p_y}{\partial x}=\frac{\partial p_y}{\partial
y}-i\frac{\partial p_x}{\partial
y}=\frac{1}{2}\left(\Gamma+i\Omega\right).
\label{firstderiQMFVorDiv}
\end{equation}
Thus, the real and imaginary parts of the QMF first derivative
determine the divergence and vorticity, respectively.  Moreover,
the complex quantum potential in Eq.~(\ref{TDQHJE}) can be expressed
in terms of divergence and vorticity by \cite{Chou9}
\begin{equation}
Q(z,t)=\frac{\hbar}{2mi}\frac{\partial p}{\partial
z}=\frac{\hbar}{4m}\left(\Omega-i\Gamma\right). \label{CQP}
\end{equation}

Additionally, we can also evaluate both the PVF divergence and
vorticity, which are given by
\begin{align}
\Gamma_{\bf P}&= \nabla \cdot {\bf P}=
\frac{\partial
p_x}{\partial x}-\frac{\partial p_y}{\partial y}=0, \\
\Omega_{\bf P}&=\left|\nabla \times {\bf P}\,\right|=
-\frac{\partial p_y}{\partial x}-\frac{\partial p_x}{\partial y}=0,
\label{VortPVF}
\end{align}
respectively, where the Cauchy-Riemann equations for the QMF have been
used.
The vanishing divergence and vorticity indicate that the PVF associated
with the QMF describes an {\it incompressible} and {\it irrotational}
flow in the complex plane except at nodes of the wave function.
Actually, this result follows from the fact that the PVF of a complex
function $f(z)$ is divergence-free and curl-free if $f(z)$ is analytic,
and vice versa \cite{Needham}.

The quantized circulation integral around quantum vortices originates
from the work term of the PVF in Eq.~(\ref{CIQMF}) \cite{Chou6},
\begin{equation}
 \gamma=\oint_C p(z) dz = \oint_C {\bf P} \cdot d \ell = 2\pi n\hbar .
\end{equation}
The PVF near a pole is expressed in terms of polar coordinates by
${\bf P}=(\gamma/2\pi r)\hat{e}_\theta$, where $\gamma$ is
the circulation and $r$ is the radial distance from the center of
the vortex.
From Eq.~(\ref{VortPVF}), the PVF vorticity is zero everywhere
{\it except at poles}.
The PVF velocity near a pole varies inversely as the distance $r$ from
the core of the vortex.
The circulation integral along a closed path enclosing the vortex is
equal to $\gamma=2\pi n\hbar$, which is independent of $r$.
Although the quantum fluid described by the streamlines around a
pole moves along a circular path, its vorticity is zero. These
features indicate that the quantum vortex described by the PVF is a
free or irrotational vortex \cite{Tritton}.


\subsection{\label{sec:DivVorPole} Divergence and vorticity around a
pole}

For a wave function with an $n$-th order node at $z=z_p$, $\psi(z) =
(z-z_p)^n f(z)$, we can evaluate the QMF first derivative, which yields
\begin{equation}
 \frac{\partial p}{\partial z} = \frac{ni\hbar}{\left(z-z_p\right)^2}
  + \frac{\partial p_s}{\partial z} ,
 \label{firstderiQMF}
\end{equation}
where $p_s(z)$ is the smooth part of the QMF.
Thus, the QMF first derivative can be approximated by the first term in
the vicinity of a pole.
For simplicity, we move the origin to the pole.
Separating the first term in Eq.~(\ref{firstderiQMF}) into its real and
imaginary parts, we obtain the divergence and vorticity around the pole
through Eq.~(\ref{firstderiQMFVorDiv}),
\begin{align}
 \Gamma& = n\hbar \ \! \frac{4xy}{(x^2+y^2)^2}, \label{AppDiv}\\
 \Omega& = n\hbar \ \! \frac{2(x^2-y^2)}{(x^2+y^2)^2} ,
\label{AppVor}
\end{align}
respectively.
These approximate forms describe the local behavior of the divergence
and vorticity in the vicinity of the pole.

\begin{figure}[t]
 \begin{center}
  \includegraphics[width=5cm]{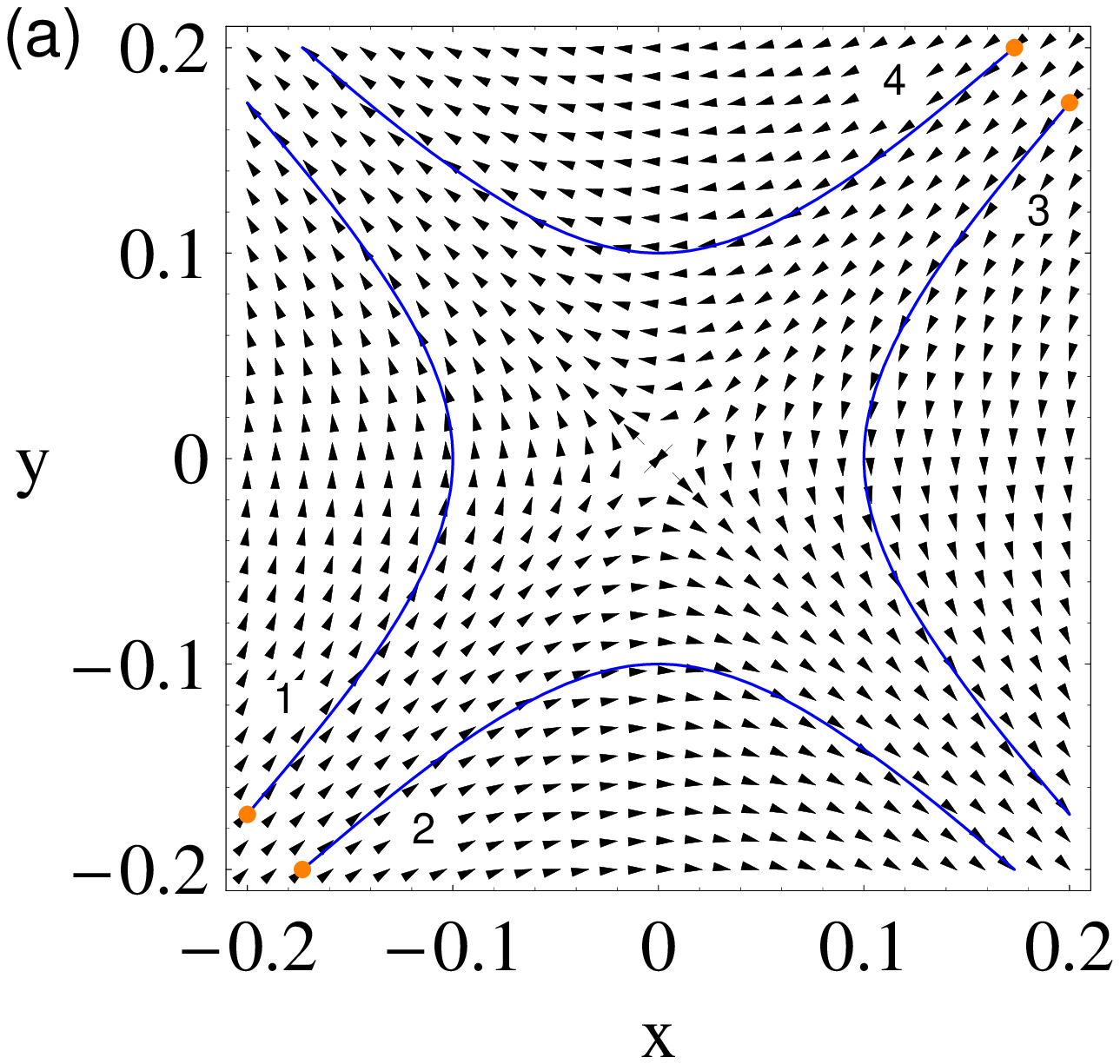}
  \includegraphics[width=5cm]{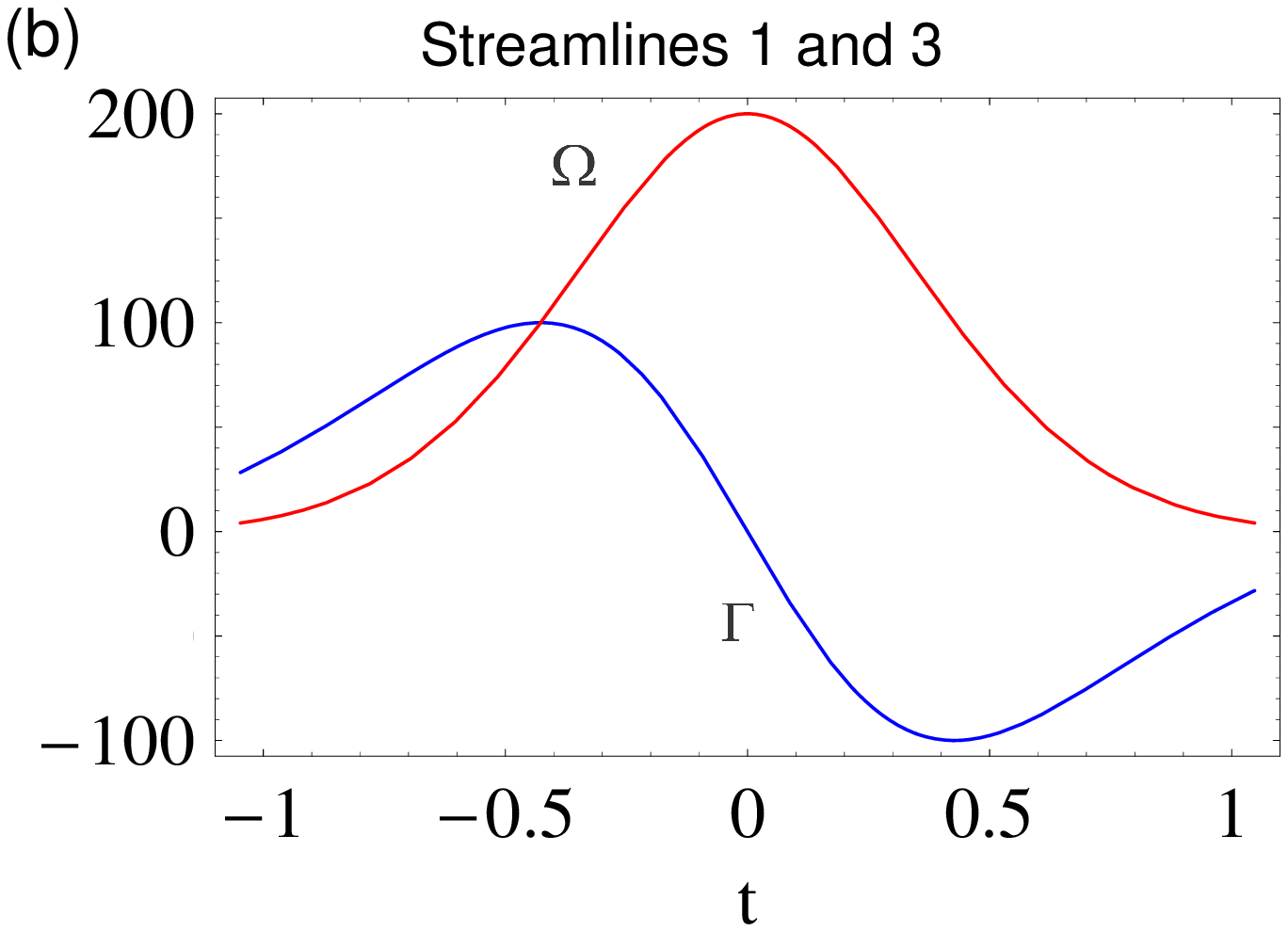}
  \includegraphics[width=5cm]{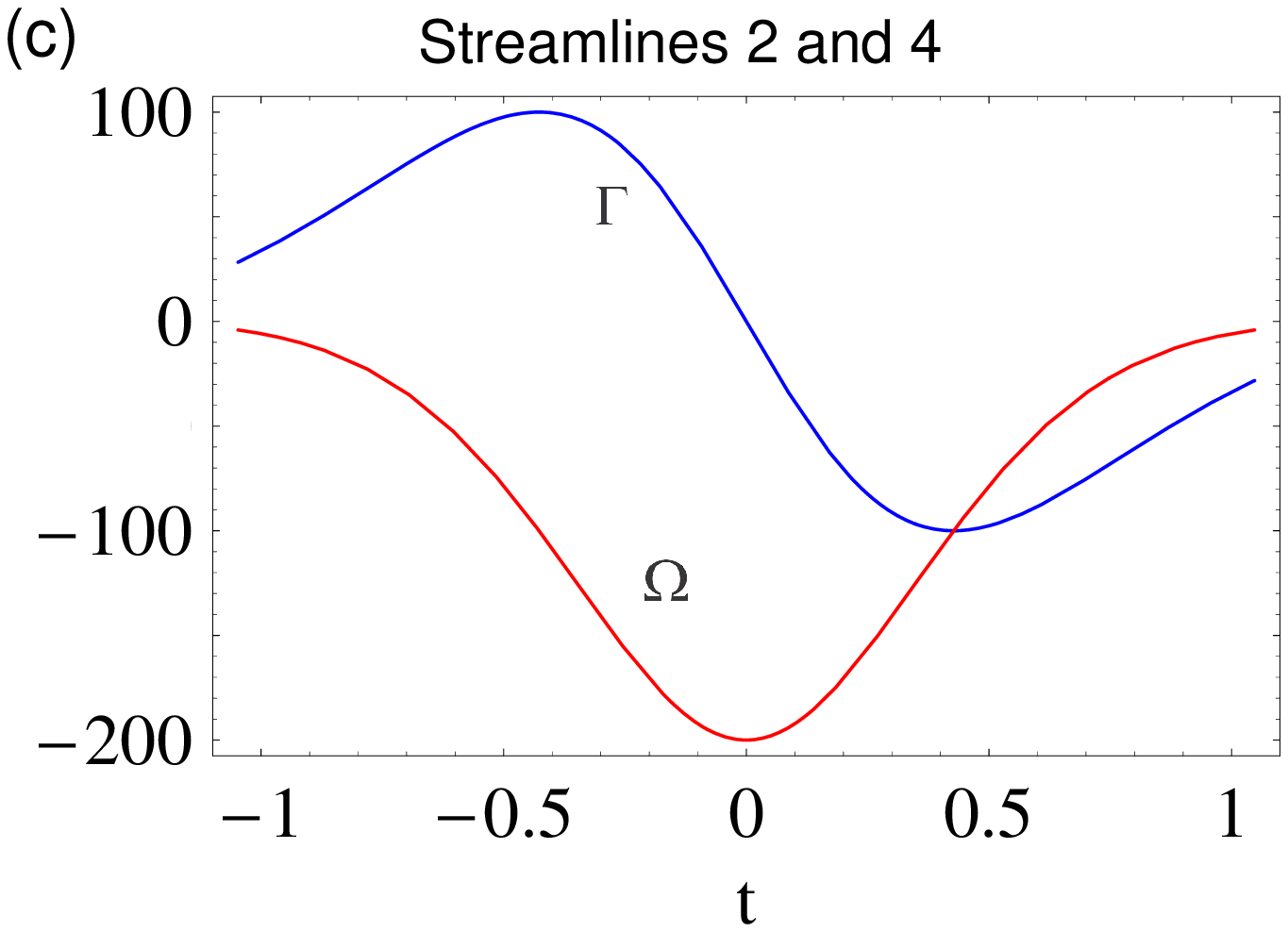}
  \caption{\label{fig2} (Color online)
   (a) The QMF displays hyperbolic flow around a pole.
   Particles start at the initial positions shown as dots and travel
   along streamlines from $t=-\pi/3$ to $t=\pi/3$.
   Variations of the QMF divergence and vorticity along:
   (b) streamlines 1 and 3 and (c) streamlines 2 and 4.}
 \end{center}
\end{figure}

Figure~\ref{fig2} presents the variations of both the QMF divergence
and vorticity along the approximate streamlines in the vicinity of a
pole. From now on, $\hbar = m = 1$ and atomic units will be used
throughout this work. In this case, we consider a wave function with
one first-order node at the origin $(n=1)$. As discussed in
Sec.~\ref{sec:LSQMF}, in Fig.~\ref{fig2}(a) we observe how the QMF
displays a hyperbolic flow around the pole.  In addition,
streamlines 1 and 3 are parametrized by $(x = \mp 0.1 \sec t, y =
\pm 0.1 \tan t)$ and streamlines 2 and 4 are parametrized by $(x =
\pm 0.1 \tan t, y = \mp 0.1 \sec t)$. In Figs.~\ref{fig2}(b) and
\ref{fig2}(c), we show the variations of both the QMF divergence and
vorticity given by Eqs.~(\ref{AppDiv}) and (\ref{AppVor}),
respectively, along the streamlines shown in Fig.~\ref{fig2}(a) from
$t=-\pi/3$ to $t=\pi/3$. As can be noticed in these figures, the
positive QMF divergence describes the local expansion of the quantum
fluid when particles approach the pole. When particles approach
turning points near the pole, the QMF divergence vanishes; when
particles leave the pole, the negative QMF divergence indicates the
local contraction of the quantum fluid. On the other hand, the QMF
vorticity describes the local rotation of the quantum fluid. During
the whole process, it can be noticed that streamlines 1 and 3
display local counterclockwise rotation, while streamlines 2 and 4
display local clockwise rotation. The QMF vorticity attains the
extrema at the turning points. Therefore, when particles approach
the pole, they rebound as they experience a repulsive force.

\section{\label{sec:QI} Quantum interference}


\subsection{\label{sec:Head}Head-on collision of two Gaussian wave
packets}


\subsubsection{On the real axis}

We consider the Gaussian wave-packet head-on collision which,
despite its simplicity, can be considered as representative of other
more complicated, realistic processes characterized by interference
(e.g., scattering problems, diffraction by slits, etc.). This
process can be described by the following total wave function
\begin{equation}
 \Psi (x,t) = \psi_L (x,t) + \psi_R (x,t) .
 \label{int1}
\end{equation}
Each wave packet ($L$ or $R$, left or right, respectively) is
represented by a free Gaussian function as
\begin{equation}
 \psi (x,t) = A_t \; e^{- (x-x_t)^2/4\tilde{\sigma}_t\sigma_0
  + i p (x - x_t)/\hbar + i E t/\hbar} ,
 \label{int2}
\end{equation}
where, for each component, $A_t = (2\pi\tilde{\sigma}_t^2)^{-1/4}$
and the complex time-dependent spreading is
\begin{equation}
 \tilde{\sigma}_t =
  \sigma_0 \left( 1 + \frac{i\hbar t}{2m\sigma_0^2} \right) .
 \label{int3}
\end{equation}
with the initial spreading $\sigma_0$. From Eq.~(\ref{int3}), the
spreading of this wave packet at time $t$ is
\begin{equation}
 \sigma_t = |\tilde{\sigma}_t| =
  \sigma_0 \sqrt{1 + \left( \frac{\hbar t}{2m\sigma_0^2} \right)^2 } .
 \label{int4}
\end{equation}
Due to the free motion, $x_t = x_0 + v_p t$ ($v_p = p/m$ is the
propagation velocity) and $E = p^2/2m$, i.e., the centroid of the
wave packet moves along a classical rectilinear trajectory. This
does not mean, however, that the average value of the wave-packet
energy is equal to $E$ since, due to the quantum potential, the
average energy is given by $\bar{E} = p^2/2m + \hbar^2/8m\sigma_0^2$
\cite{ASSanz9}.  We observe two contributions in this expression.
The former is associated with the translation of the wave packet
traveling at velocity $v_p$, while the latter is related to its
spreading at velocity $v_s$ and has, therefore, a purely
quantum-mechanical origin.  The {\it effective} momentum $p_s$ can
then be written as $p_s=\hbar/2\sigma_0$, which resembles
Heisenberg's uncertainty relation.

The relationship between $v_p$ and $v_s$ plays an important role in
effects that can be observed when dealing with wave packet
superpositions \cite{ASSanz9}. Defining the timescale $\tau =
2m\sigma_0^2/\hbar$, we note that, if $t \ll \tau$, the width of the
wave packet remains basically constant with time, $\sigma_t \approx
\sigma_0$ (i.e., for practical purposes, it is roughly
time-independent up to time $t$). This condition is equivalent to
having an initial wave packet prepared with $v_s \ll v_p$. Thus, the
translational motion will be much faster than the spreading of the
wave packet.  On the contrary, if $t \gg \tau$, which is equivalent
to $v_s \gg v_p$, the width of the wave packet increases linearly
with time ($\sigma_t \approx \hbar t/2m\sigma_0$), and the wave
packet spreads very rapidly in comparison with its advance along
$x$.  Of course, in between, there is a smooth transition; from
Eq.~(\ref{int4}), it is shown that the progressive increase of
$\sigma_t$ describes a hyperbola when this magnitude is plotted {\it
vs} time. Thus, we can control in the interference process the
spreading and translational motions which determine the wave packet
dynamics \cite{ASSanz8,ASSanz9}.


\subsubsection{In the complex plane}

In conventional quantum mechanics, the interference pattern
transiently observed on the real axis is attributed to the
constructive and destructive interference between two
counter-propagating components of the total wave function in
Eq.~(\ref{int1}).  In contrast, in the framework of the complex
quantum Hamilton-Jacobi formalism, the total wave function is
analytically continued to the complex plane from $\Psi(x,t)$ to
$\Psi(z,t)$.  Therefore, two propagating wave packets \emph{always
interfere with each other in the complex plane}, and this leads to a
persistent pattern (line) of nodes and stagnation points which
rotates counterclockwise with time \cite{ASSanz8,Chou9}.

The nodal positions in the complex plane can be determined analytically
by solving the equation $\Psi(z,t)=0$, resulting
\begin{equation}
z_n(t)=\frac{i\pi\left(n+1/2\right)}{\left[imv_p/\hbar-\left(x_0-v_pt\right)
/\left(2\sigma_0\tilde{\sigma}_t\right)\right]},
\end{equation}
where $n = 0, \pm1, \pm2, \ldots$.  Here, we have assumed
$x_R=-x_L=x_0$ and $v_{pL}=v_{pR}=v_p$.  Splitting this expression
into its real and imaginary parts, i.e., $z_n(t)=x_n(t)+iy_n(t)$, we
obtain
\begin{align}
x_n(t)&=\pi\left(n+\frac{1}{2}\right)\frac{\hbar}{m}
\left[\frac{x_0t+v_p\tau^2}{x_0^2+v_p^2\tau^2}\right], \label{nodeX} \\
y_n(t)&=\pi\left(n+\frac{1}{2}\right)\left[
\frac{2\sigma_0^2\left(v_pt-x_0\right)}{x_0^2+v_p^2\tau^2}\right],
\label{nodeY}
\end{align}
respectively, where $\tau = 2m\sigma_0^2/\hbar$ is the timescale for
the Gaussian wave packet. In addition, dividing Eq.~(\ref{nodeY}) by
Eq.~(\ref{nodeX}) yields the analytical expression for the
(time-dependent) polar angle describing the angular position of the
nodal line with respect to the positive real axis,
\begin{equation}
 \theta(t) = (\tan)^{-1}\left[\frac{y_n(t)}{x_n(t)}\right]
  = (\tan)^{-1} \left[ \frac{\tau\left(v_pt-x_0\right)}{x_0t+v_p\tau^2}
   \right] ,
 \label{nodalAngle}
\end{equation}
which does not depend on $n$.  From this expression, we can
calculate the rotation rate of the nodal line in the complex plane,
\begin{equation}
\omega(t)=\frac{d\theta(t)}{dt}=\frac{\hbar}{2m\sigma_t^2} ,
\label{NodalRate}
\end{equation}
where $\sigma_t$ is given in Eq.~(\ref{int4}). This equation
indicates that the rotation rate is completely determined by the
initial spreading of the Gaussian wave packet $\sigma_0$.  In
addition, this rate is always positive and decays monotonically to
zero as $t$ goes to $\infty$. From Eqs.~(\ref{nodeX}) and
(\ref{nodeY}), we can also determine the node separation distance
between two consecutive nodes
\begin{align}
 d(t) & = \sqrt{ [x_{n+1}(t)-x_n(t)]^2 + [y_{n+1}(t) - y_n(t)]^2}
  \nonumber\\
 & = \frac{\pi\hbar\sigma_t}{p_s\sqrt{x_0^2+v_p^2\tau^2}},
 \label{NodalDist}
\end{align}
where $p_s = \hbar/2\sigma_0$ is the effective momentum.
This distance is independent of $n$ and it increases with time.
Moreover, eliminating $t$ in Eqs.~(\ref{nodeX}) and (\ref{nodeY})
yields the $n$th node trajectory describing the time evolution of the
node given by
\begin{equation}
y_n = \left(\frac{v_p\tau}{x_0}\right)x_n -
(2n+1)\left(\frac{\pi\sigma_0^2}{x_0}\right). \label{NodalTraj}
\end{equation}
Consequently, these nodal trajectories with the same slope and
different intercepts are parallel to each other.

When $t=0$, the initial angle of the nodal line is $\theta_0 =
(\tan)^{-1}(-x_0/v_p\tau)$.  As described by Eq.~(\ref{NodalRate}), the
positive rotational rate indicates that the polar angle of the nodal
line increases monotonically with time.
The nodal line thus rotates counterclockwise from the initial angle to
a maximum or limiting value $\theta_\infty = (\tan)^{-1}(v_p\tau/x_0)$
when $t \to \infty$.
The angular displacement from $t=0$ to $t=\infty$ is always equal
to $\Delta \theta = \theta_\infty - \theta_0 = \pi/2$, because the
product of the slopes of the nodal lines is equal to $(\tan \theta_0)
(\tan \theta_\infty) = -1$.
In particular, if both wave packets are initially very far apart (i.e.,
$x_0 \to \infty$), but move with a finite velocity $v$, or they are at
an arbitrary finite distance, but $v = 0$, the nodal line ends up
aligned with the real axis.
Otherwise, the nodal line starts at some angle $\theta_0$ and then
evolves with the angular displacement $\Delta\theta=\pi/2$ until it
reaches the limit angle $\theta_\infty$.
In addition, the initial nodal line is perpendicular to all nodal
trajectories in Eq.~(\ref{NodalTraj}).
Then, the nodal line rotates counterclockwise with time and it becomes
parallel to nodal trajectories when $t$ approaches infinity.

Interference features are observed on the real axis only when the
nodal line is near the real axis. When the nodal line coincides with
the real axis, the maximum interference features can be observed in
real space.  At this time, setting $y_n = 0$ in
Eq.~(\ref{NodalTraj}), we recover the expression for the positions
of nodes on the real axis, $x_n = (n+1/2)\lambda/2$ where $\lambda =
2\pi\hbar/mv_p$.  During the evolution of the nodal line, its intersections
with nodal trajectories determine the positions of the nodes.
As can be noticed, the time evolution of the nodal line in the complex
plane therefore provides an elegant interpretation of quantum
interference.


\subsection{Case 1: $v_p > v_s$}

We first consider the case where the relative propagation velocity
is larger than the spreading rate of the wave packets.  The
following initial conditions for Gaussian wave packets are used:
$x_{0L} = - 10 = - x_{0R}$, $v_{pL} = 2 = -v_{pR}$ and $\sigma_0 =
\sqrt{2}$. With these conditions, maximal interference occurs at
$t=5$ on the real axis and the propagation and spreading velocities
are given by $v_p=2$ and $v_s=\sqrt{2}/4$, respectively.

\begin{figure}[t]
 \begin{center}
 \includegraphics[width=6.5cm]{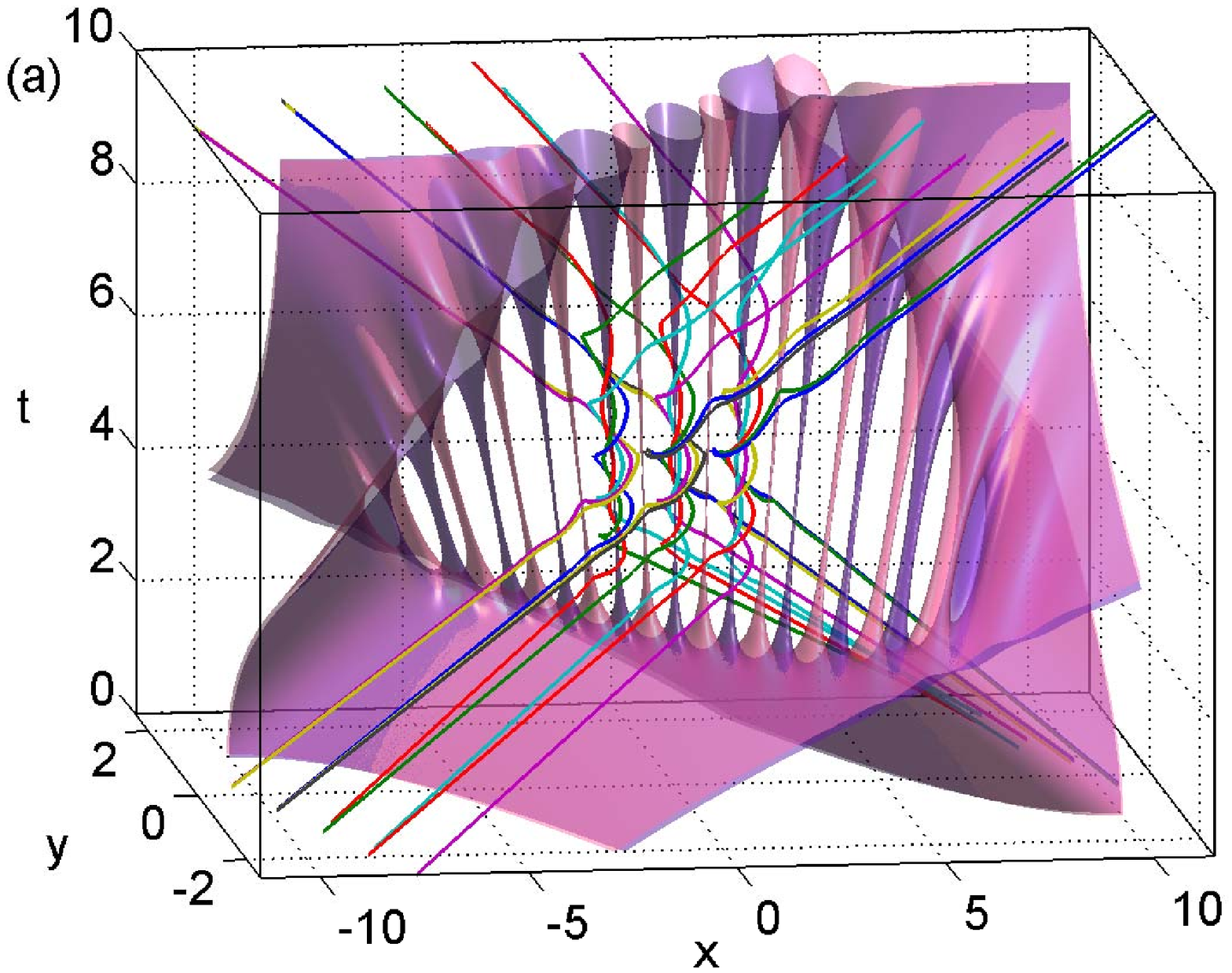}
 \includegraphics[width=6.5cm]{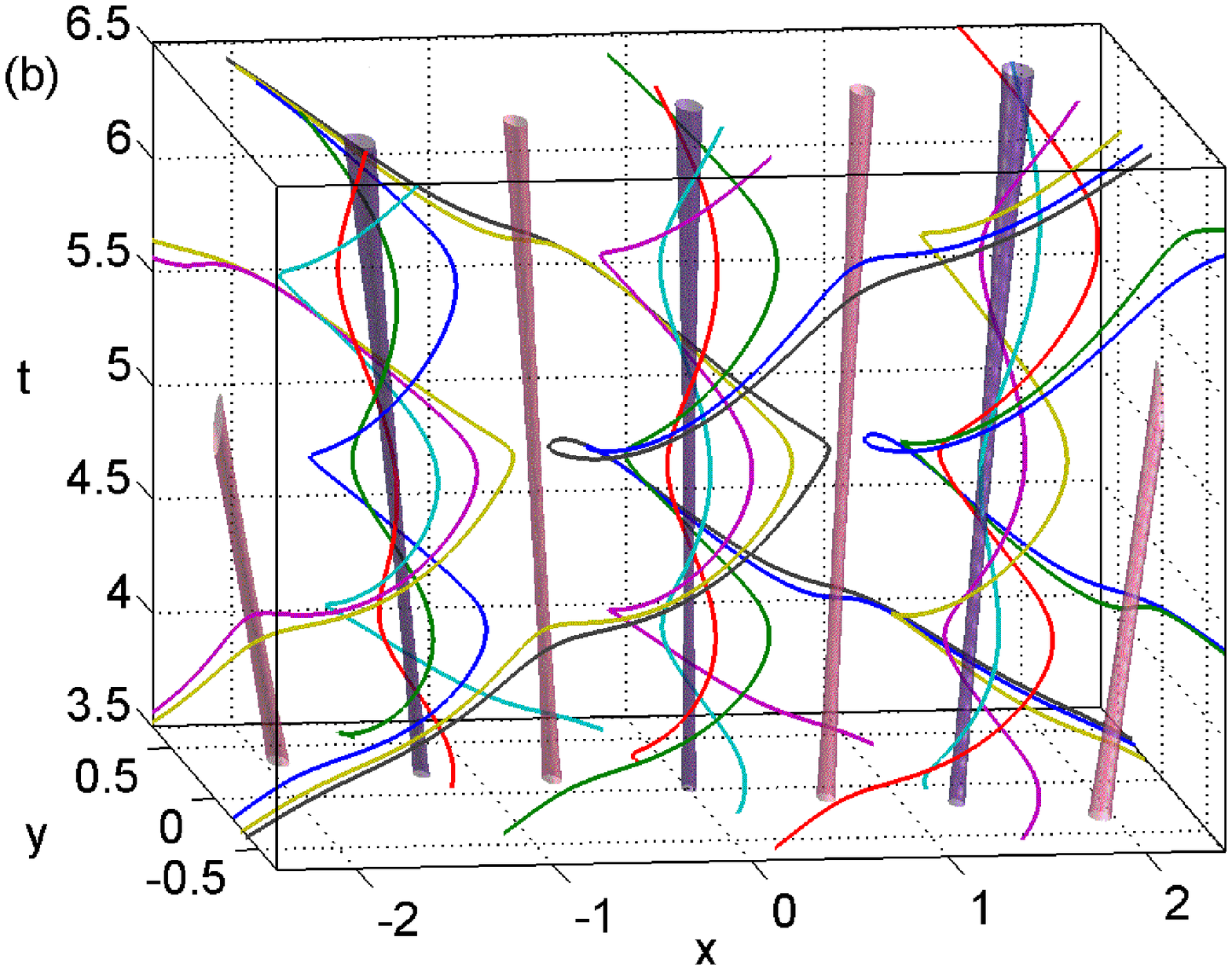}
 \caption{\label{fig3} (Color online)
  (a) Quantum caves and complex quantum trajectories for the head-on
  collision of two Gaussian wave packets with the relative velocity
  larger than the spreading rate.  These trajectories launched
  from the isochrone arrive at the real axis at $t=5$.
  The caves are formed with the isosurfaces $|\Psi (z,t)|=0.053$
  (pink/lighter gray surface) and $|\partial \Psi (z,t)/ \partial z|=0.106$
  (violet/darker gray surface).
  (b) Complex quantum trajectories displaying
  helical wrapping around the stagnation tubes (violet/darker gray) and hyperbolic
  deflection around the vortical tubes (pink/lighter gray).}
 \end{center}
\end{figure}


\subsubsection{Quantum caves with quantum trajectories}

Figure \ref{fig3}(a) displays complex quantum trajectories with the
quantum caves consisting of the isosurfaces $|\Psi (z,t)| = 0.053$
and $|\partial \Psi (z,t)/\partial z|= 0.106$ from $t=0$ to $t=10$
in a time-dependent three-dimensional Argand plot. As discussed in
Sec.~\ref{sec:LSQMF}, the QMF local structures near
stagnation points and poles provide a qualitative description of the
behavior of these trajectories. It is clearly seen that stagnation
and vortical tubes alternate with each other, and the centers of the
tubes are stagnation and vortical curves. Trajectories display
helical wrapping around the stagnation tubes and they are deflected
by the vortical tubes to show hyperbolic indentations in three
dimensional space, as shown in Fig.~\ref{fig3}(b).  These
trajectories, which display complicated paths around stagnation and
vortical tubes, depict how probability flows.  In addition,
trajectories from different launch points wrap around the same
stagnation curve and remain trapped for a certain time interval.
Then, they separate from the stagnation curve.  This
counterclockwise circulation of trajectories can be viewed as a
resonance process.  This phenomenon characterizes long-range
correlation among trajectories arising from different starting
points.  Trajectories may wrap around the same stagnation curve with
different wrapping times and numbers of loops. Trajectories starting
from the isochrone with the small initial separation may wrap around
different stagnation curves and then end up with large separations.
In this way, interference leads to the formation of quantum caves and
the topological structure displayed by complex quantum trajectories.

\begin{figure}[t]
 \begin{center}
 \includegraphics[width=6.5cm]{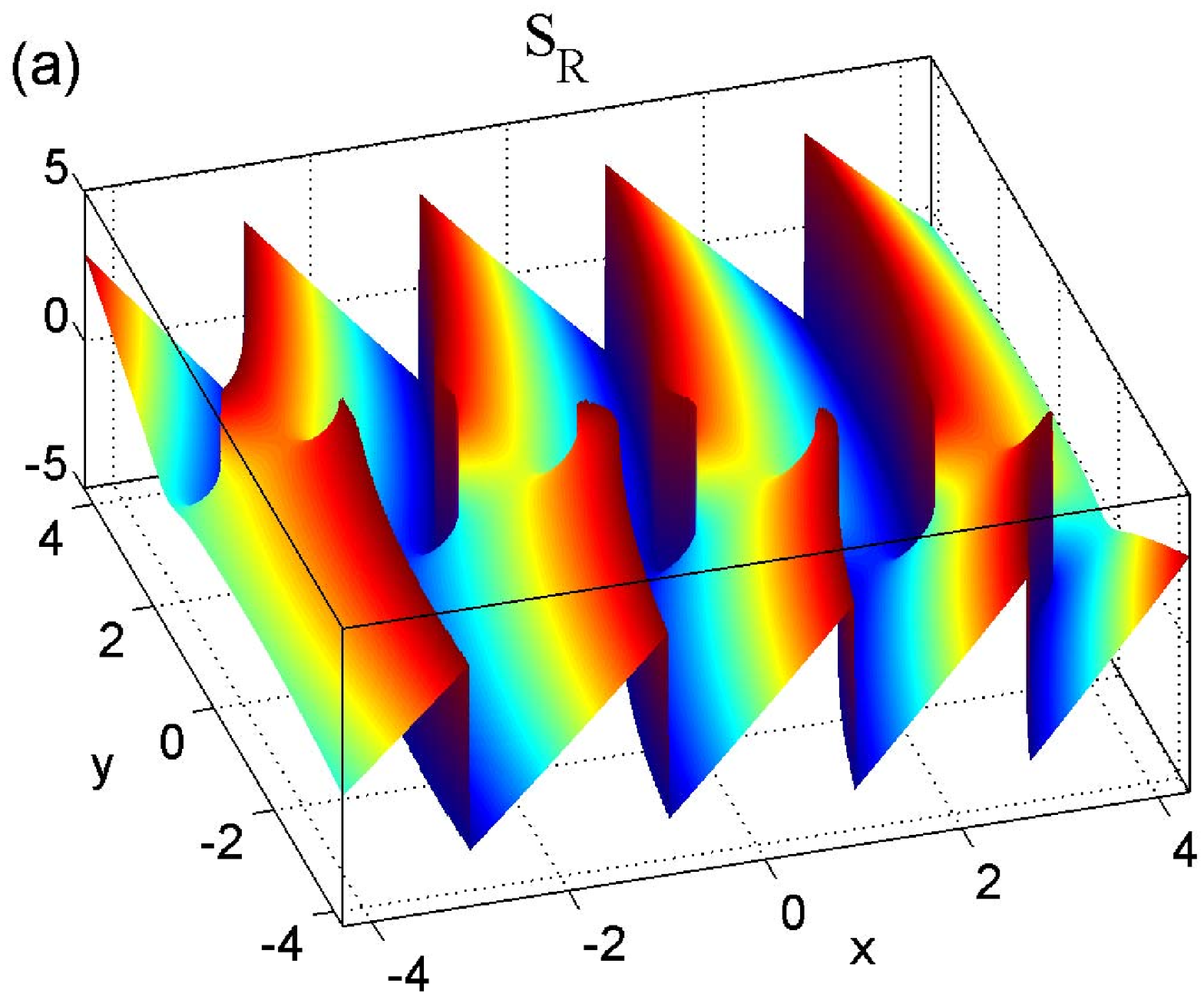}
 \includegraphics[width=6.5cm]{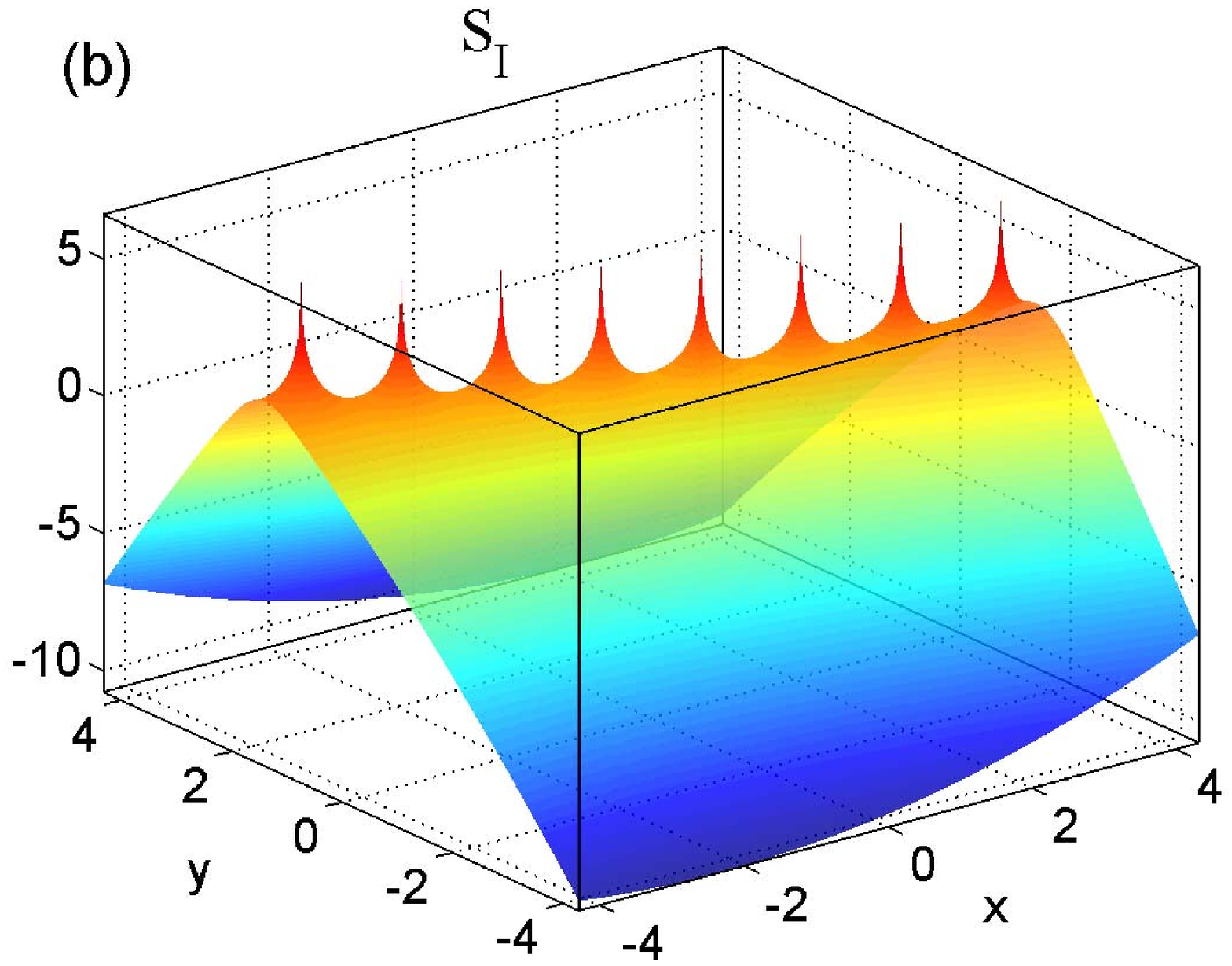}
 \caption{\label{fig4} (Color online)
  Real (a) and imaginary (b) part of the complex
  action for the complex-extended wave function associated to
  Eq.~(\ref{int1}) at $t=2.5$.
  For the real part, the principal value of the multivalued phase
  function is shown.
  Blue to red degradation indicates the transition from lower to higher
  values of the corresponding functions (in real part of the complex action, the range
  goes from $-\pi$ to $\pi$).}
 \end{center}
\end{figure}


\subsubsection{Dislocations of the complex action}

The complex action function $S(z,t)$ displays fascinating features in
the complex plane.
Decomposing this function into its real and imaginary parts
$S=S_R+iS_I$, we write the wave function as $\psi(z,t)=\exp(-S_I/\hbar)
\exp(iS_R/\hbar)$.
According to this expression, the real and imaginary parts of the
complex action determine the {\it phase} and {\it amplitude} of the
wave function, respectively.
Figure~\ref{fig4} displays the real and imaginary parts of the complex
action for the complex-extended wave function in Eq.~(\ref{int1}) at
$t=2.5$.
Figure~\ref{fig4}(a) displays the principal zone of the phase of the
wave function in the range $-\pi \leq \arg(S_R) \leq \pi$.
Figure~\ref{fig4}(b) displays the imaginary part $S_I$ of the
complex action in the complex plane.  The vanishing of the wave
function at nodes indicates that the imaginary part of the complex
action tends to positive infinity at nodes.  The peaks in
Fig.~\ref{fig4}(b) correspond to the nodes in the wave function.

The quantized circulation integral around a node in the wave function
can be related to the change in the phase of the wave function and this
integral can be expressed in terms of the PVF by \cite{Chou6}
\begin{equation}
 \oint_C p(z)dz=\oint_C {\bf P} \cdot d \ell = 2\pi n\hbar
  = \Delta_C S_R ,
 \label{QISCI}
\end{equation}
where $C$ denotes a simple closed curve in the complex plane.  The
quantized circulation integral around a node of the wave function
originates from the discontinuity in its phase.  The PVF displays
\emph{counterclockwise circular flow} in the vicinity of a node
\cite{Chou6,Chou7}. Nodes in the complex-extended wave function in
Eq.~(\ref{int1}) are first-order nodes $(n=1)$.  As shown in
Fig.~\ref{fig4}(a), if we travel around a first-order node
counterclockwise along a closed path, it follows from
Eq.~(\ref{QISCI}) that the phase displays a sharp discontinuity of
$2\pi$.  Actually, the phase of the wave function is a multivalued
function in the complex plane.  Thus, if we travel counterclockwise
around a node, we then go through the branch cut from one Riemann
sheet to another.  Through a continuous closed circuit around a node
continuing on all the sheet, the phase function generates a helicoid
along the vertical axis.  Analogous to the case in Bohmian
mechanics, phase singularities at nodes in the complex plane can be
interpreted as \emph{wave dislocations} \cite{NyeBerry,HollandBook}.

\begin{figure}[t]
 \begin{center}
 \includegraphics[width=6cm]{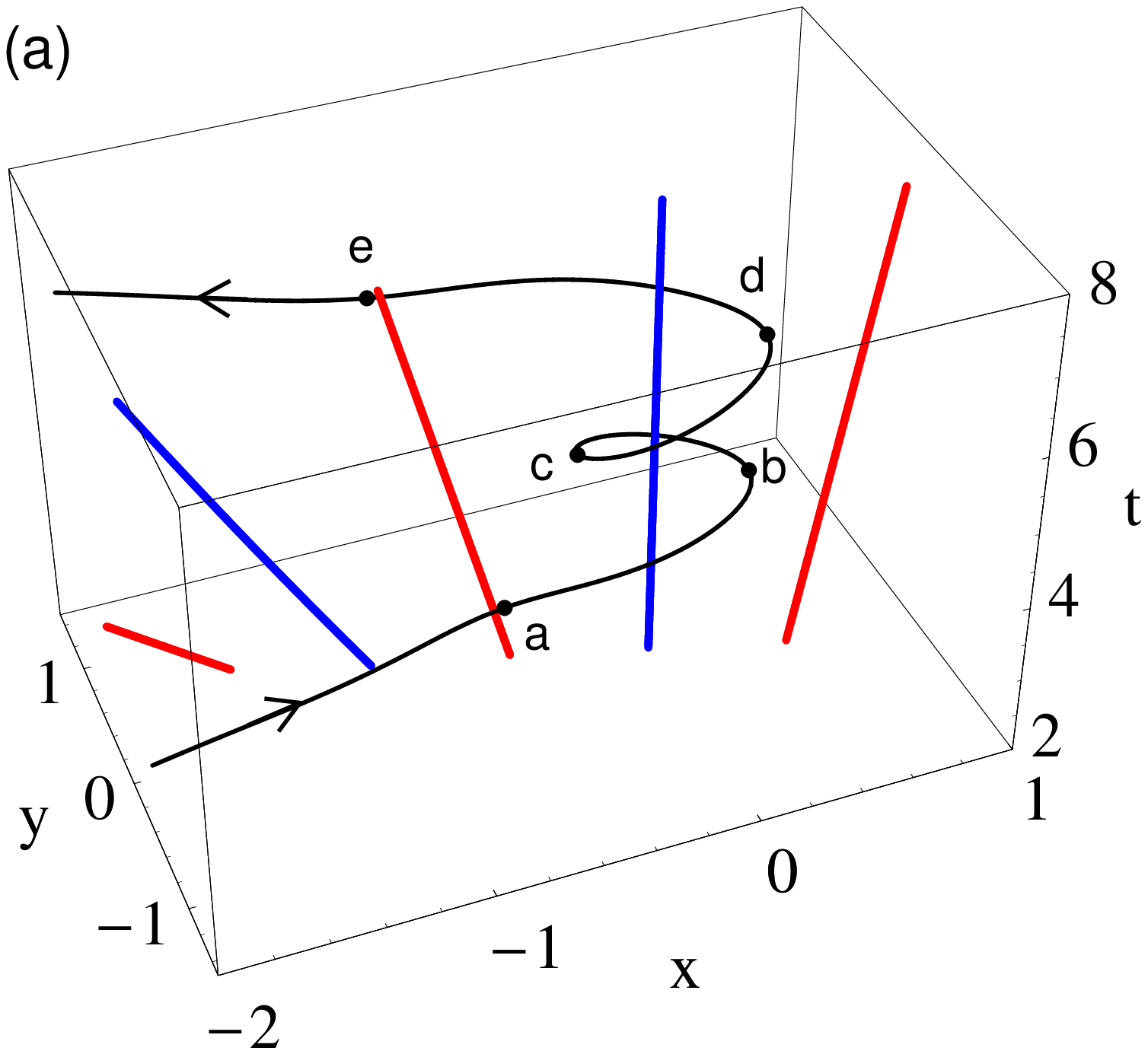}
 \includegraphics[width=6.5cm]{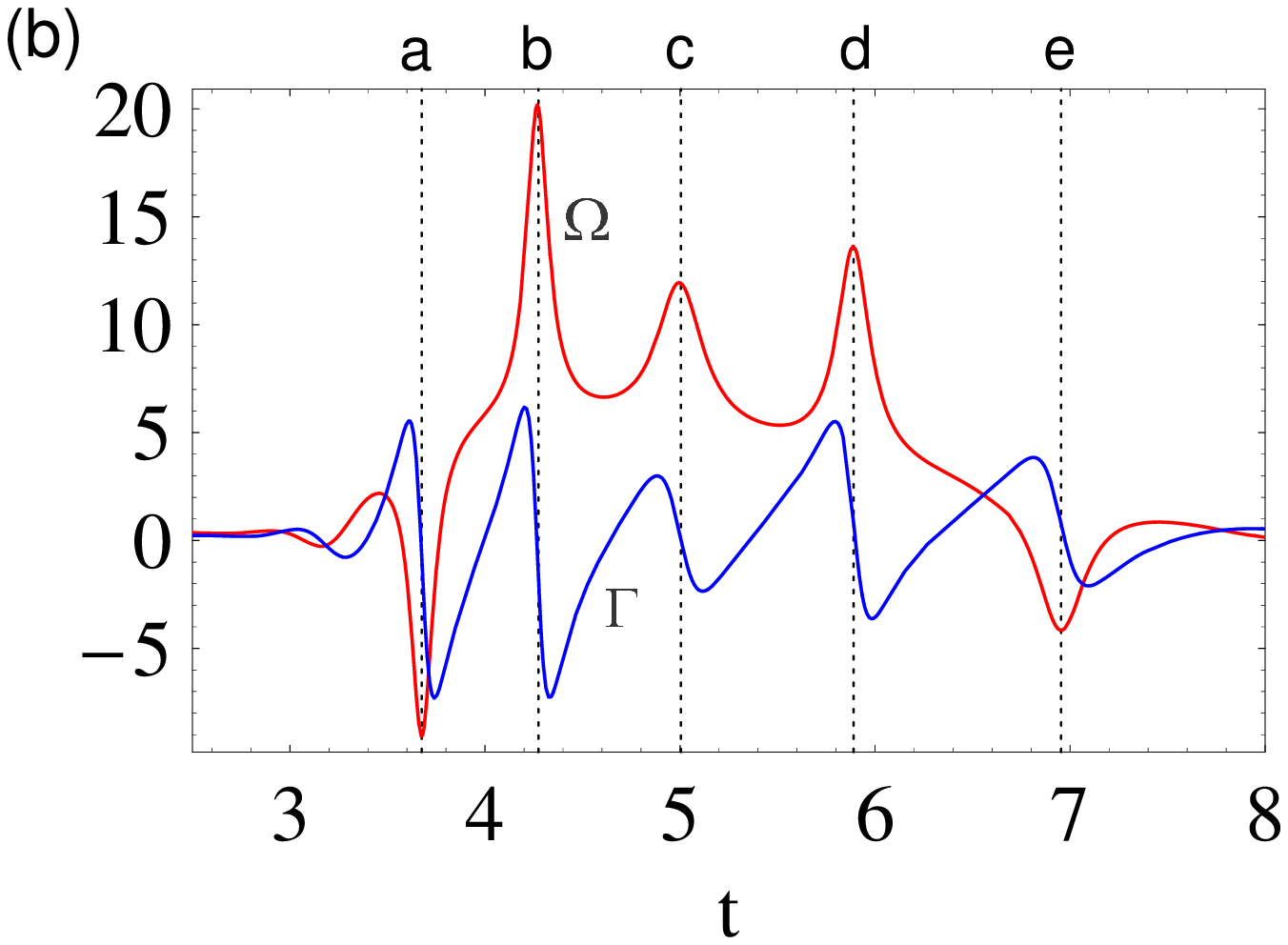}
 \caption{\label{fig5} (Color online)
  (a) Complex quantum trajectory launched from $z=-9.11016-1.17309i$
  and reaching the real axis at $z=-0.3$ when $t=5$.
  As can be noticed, this trajectory undergoes a helical wrapping
  around an stagnation curve (blue/darker gray) and hyperbolic
  deflection around the vortical curves (red/light gray).
  (b) Vorticity and divergence along the trajectory.}
 \end{center}
\end{figure}


\subsubsection{QMF divergence and vorticity}

Figure~\ref{fig5}(a) shows a trajectory launched from
$z=-9.11016-1.17309i$ which later arrives at $z=-0.3$, when maximal
interference occurs at $t=5$, and Fig.~\ref{fig5}(b) presents the
divergence and vorticity along this trajectory.  When the particle
approaches a turning point, its velocity undergoes a rapid change
and the divergence and vorticity display sharp fluctuations.  When
the particle approaches the vortical curve at position $a$ along the
direction of streamline 2 shown in Fig.~\ref{fig2}(a), the
trajectory displays hyperbolic deflection and the divergence and
vorticity display analogous variations, as shown in
Fig.~\ref{fig2}(c). Around position $a$, when the particle
approaches the vortical curve, the positive divergence indicates the
local expansion of the quantum fluid.  When the particle arrives at
the turning points, the divergence vanishes.  Then, when it leaves
the vortical curve, the negative divergence indicates the local
contraction of the quantum fluid.  On the other hand, the negative
vorticity describes the local clockwise rotation of the quantum
fluid.

When the trajectory displays helical wrapping around the stagnation
curve, the particle is trapped between two vortical curves.  When
the particle approaches the turning points $b$ and $d$ along the
direction of streamline 1 shown in Fig.~\ref{fig2}(a) and approaches
the turning point $c$ along the direction of streamline 3, the
divergence and vorticity in Fig.~\ref{fig5}(b) display analogous
variations shown in Fig.~\ref{fig2}(b). When the particle approaches
and leaves the vortical curve, the divergence describes the local
expansion and contraction of the quantum fluid in the vicinity of
the vortical curve.  The positive vorticity indicates the
counterclockwise rotation of the quantum fluid.  Finally, when the
particle leaves the stagnation curve, it approaches the vortical
curve at position $e$ along the direction of streamline 4 shown in
Fig.~\ref{fig2}(a). Again, the divergence and vorticity around
position $e$ display similar fluctuations shown in
Fig.~\ref{fig2}(c).  Therefore, the variations of the divergence and
vorticity around a pole analyzed in Sec.~\ref{sec:DivVorPole}
provide a qualitative description of the local behavior of the
complex quantum trajectories in the vicinity of vortical tubes.

\begin{figure}[t]
 \begin{center}
 \includegraphics[width=6.5cm]{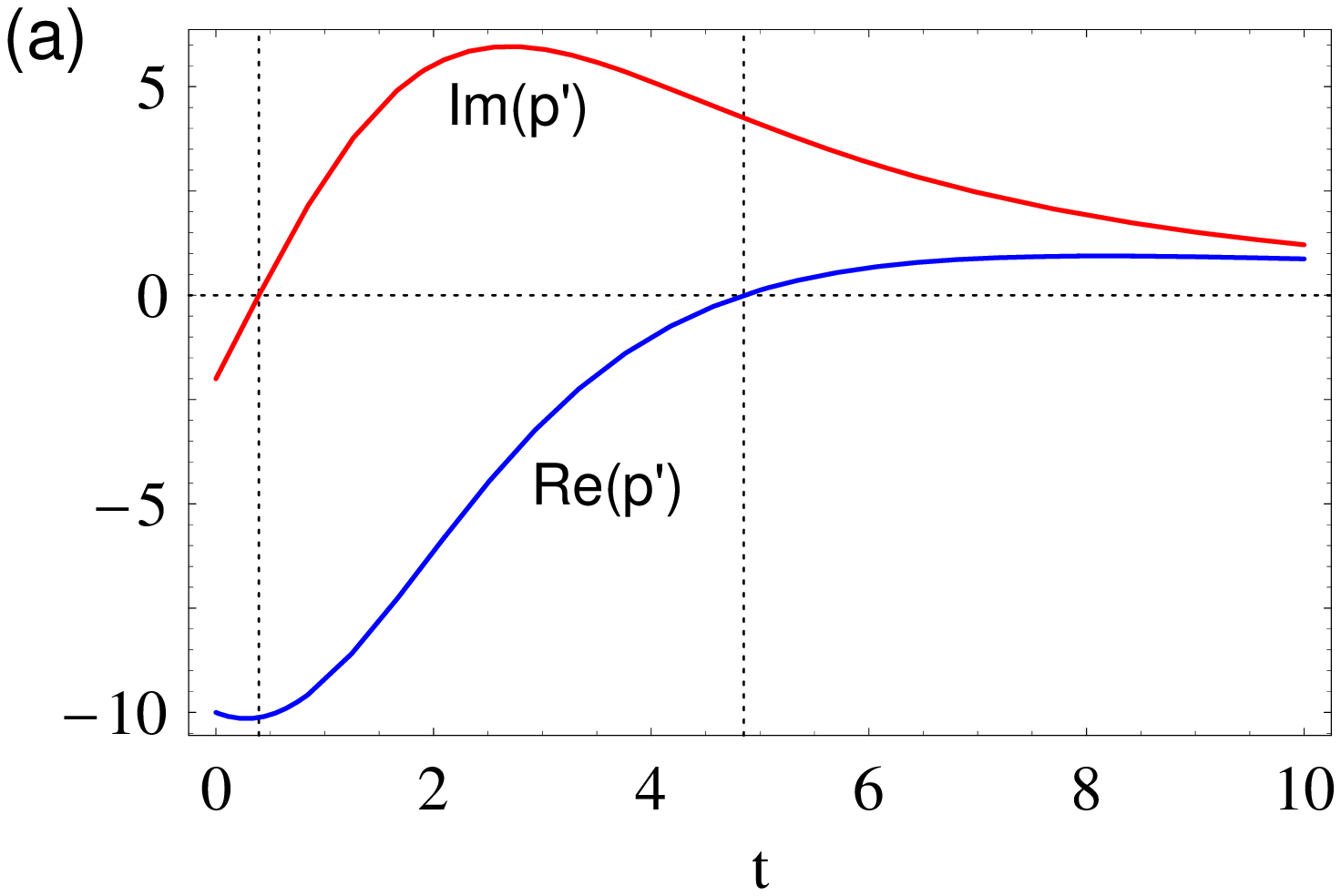}
 \includegraphics[width=6.5cm]{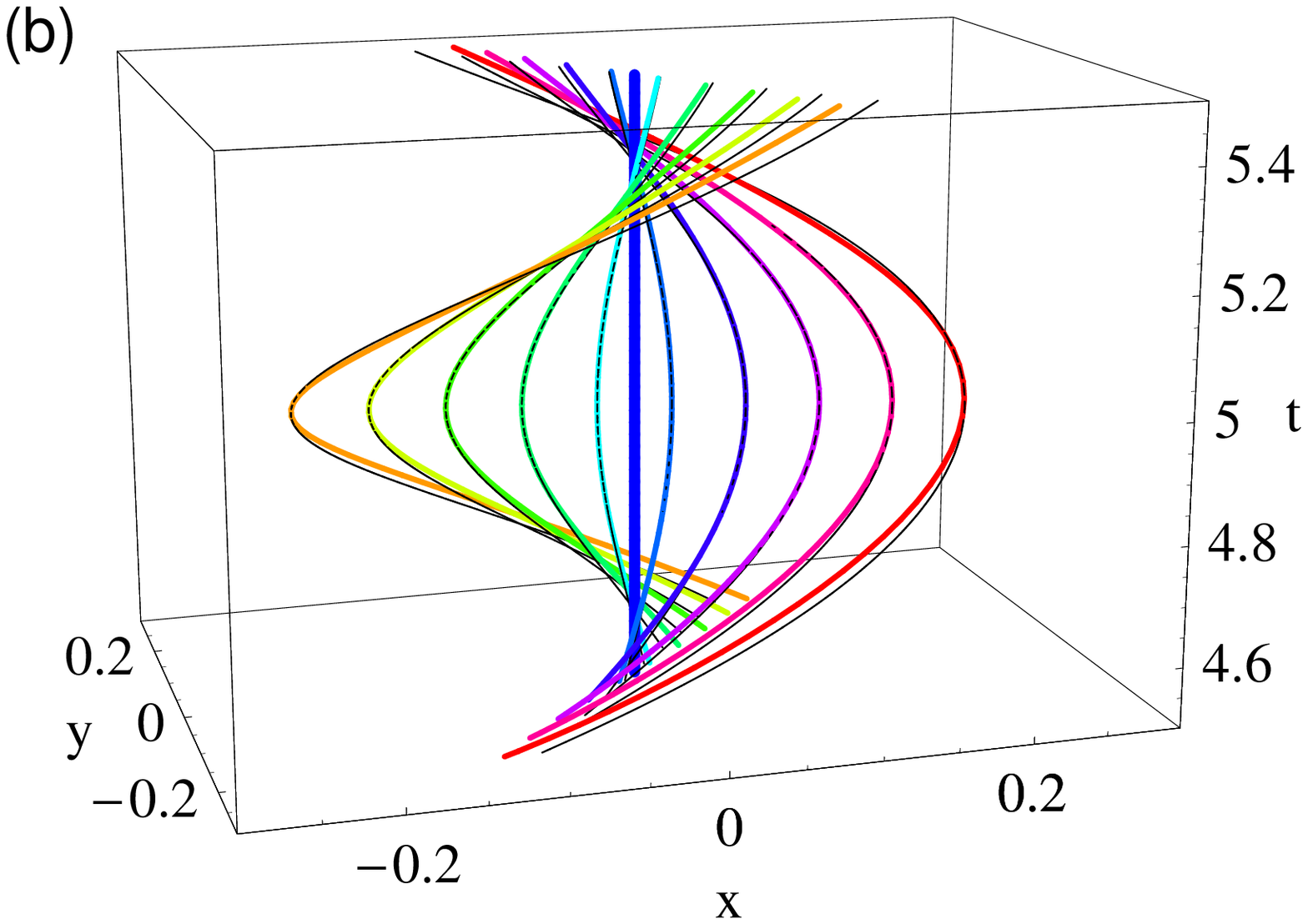}
 \caption{\label{fig6} (Color online)
  (a) QMF first derivative at $z=0$.
  (b) The exact trajectories (thick curves) starting from the isochrone
  arrive at the real axis at $t=5$ with the approximate trajectories
  (thin curves) determined by Eq.~(\ref{ThreeDfirstorderTJ}) around the
  stagnation curve along $z=0$.}
 \end{center}
\end{figure}


\subsubsection{Approximate quantum trajectories}

As discussed in Sec.~\ref{sec:LSQMF}, the QMF first derivative
evaluated at stagnation points determines the local structure around
these points \cite{Chou7}.  Since it is noted from Eq.~(\ref{int1})
that there is always constructive interference at the origin,  the
origin is a stagnation point at all times.  As an example,
Fig.~\ref{fig6}(a) presents the QMF first derivative at the origin
from $t=0$ to $t=10$.  The real part of the QMF first derivative
indicates that the QMF displays convergent flow around the
stagnation point until $t=4.851$.  Then, the real part of the QMF
first derivative becomes positive and the QMF displays divergent
flow.  On the other hand, the QMF initially displays clockwise flow
around the stagnation point. After $t=0.395$, the QMF displays
counterclockwise flow.  As displayed in Fig.~\ref{fig3},
trajectories exhibit helical wrapping around the stagnation curve
along $z=0$ from approximately $t=3.5$ to $t=6.5$. During this time
interval, the positive imaginary part of the QMF first derivative in
Fig.~\ref{fig6}(a) describes the counterclockwise flow of the
quantum fluid.  However, the real part of the QMF first derivative
changes sign at $t=4.851$. Particles initially converge to the
stagnation point and they are gradually repelled by the stagnation
point after $t=4.851$. Finally, these particles depart from the
stagnation point.  Therefore, the QMF first derivative
 at the stagnation point qualitatively explains the behavior of
trajectories in the vicinity of the stagnation point.

Additionally, Fig.~\ref{fig6}(b) shows that the trajectories
starting from the isochrone arrive at the real axis at $t=5$ with
the approximate trajectories around the stagnation point given in
Eq.~(\ref{ThreeDfirstorderTJ}).  Here, the stagnation point in
spacetime for Eq.~(\ref{ThreeDfirstorderTJ}) is chosen to be $z=0$
and $t=5$.  The positions for the approximate trajectories are set
to be the same as those for the exact trajectories at $t=5$.  As
shown in this figure, the trajectory determined by
Eq.~(\ref{ThreeDfirstorderTJ}) is a good approximation of the exact
trajectory, provided that the approximate trajectory is close to the
stagnation point in spacetime.


\subsubsection{Wrapping time}

These complicated features for trajectories arise from the complex
quantum potential in Eq.~(\ref{CQP}). Both the QMF divergence and
vorticity characterize the turbulent flow of complex quantum
trajectories. Moreover, the variation of the QMF vorticity provides
a feasible method to define the wrapping time for a specific
trajectory around a stagnation curve.  The wrapping time can be
defined by the time interval between the first and last minimum of
$\Omega$ comprising a region with the positive vorticity. Within
this time interval, the positive vorticity describes the
counterclockwise twist of the trajectory displaying the interference
dynamics, and the stagnation points and poles significantly affect
the motion of the particle.  For example, Fig.~\ref{fig5}(b)
indicates that the particle undergoes counterclockwise wrapping
around the stagnation curve from $t=3.676$ to $t=6.954$ and hence
the wrapping time is $t_W=3.278$.  Different trajectories have
different wrapping times.  Therefore, the average wrapping time can
be utilized to define the `lifetime' for the interference process
observed on the real axis.

\begin{figure}[t]
 \begin{center}
 \includegraphics[width=7.5cm]{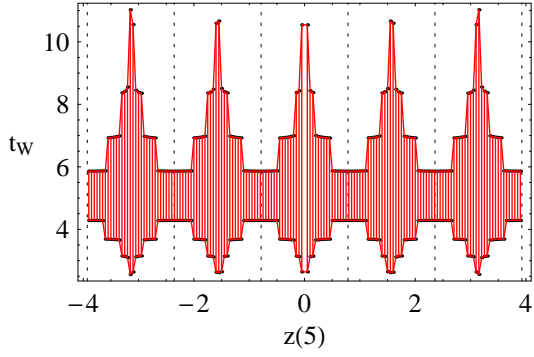}
 \caption{\label{fig7} (Color online)
  Wrapping times corresponding to the first and last minimum of the
  vorticity, which comprise a region with positive vorticity for
  trajectories launched from the isochrone.
  These trajectories arrive at the real axis at $t=5$, covering a range
  which goes from $z(5)=-3.9$ to $z(5)=3.9$ with an increment of
  $\Delta z = 0.05$.
  The vertical dotted lines denote the positions of the nodes on the
  real axis at $t=5$.}
 \end{center}
\end{figure}

In Fig.~\ref{fig7} we present the wrapping times that correspond to
the first and last minimum of the vorticity, which comprise a region
with positive vorticity for trajectories launched from the isochrone
arriving at the real axis at $t=5$. Since the process described in
Eq.~(\ref{int1}) in the complex plane is symmetric with respect to
the origin, the data shown in Fig.~\ref{fig7} display the same
symmetry.  As displayed in Fig.~\ref{fig3}(b), trajectories can wrap
around stagnation curves with different wrapping times and numbers
of loops.  Figures~\ref{fig7} and \ref{fig3}(b) indicate that
trajectories wrapping around stagnation curves with small rotational
radius are trapped between vortical curves for a long time, and this
leads to a long wrapping time for these trajectories. On the
contrary, if trajectories wrapping around stagnation curves with
large rotational radius, when they approach vortical curves, they
experience a significant repelling force provided by QMF poles.
Thus, these trajectories can easily escape capture by stagnation
curves, and this results in a short wrapping time.  In addition, it
is noted in Fig.~\ref{fig7} that all trajectories display helical
wrapping from approximately $t=4.3$ to $t=5.9$. Furthermore, the
average wrapping time for these trajectories is $\bar{t}_W=3.24$,
and it can be used to define the lifetime for the interference
process in this case.

\begin{figure}[t]
 \begin{center}
 \includegraphics[width=6.5cm]{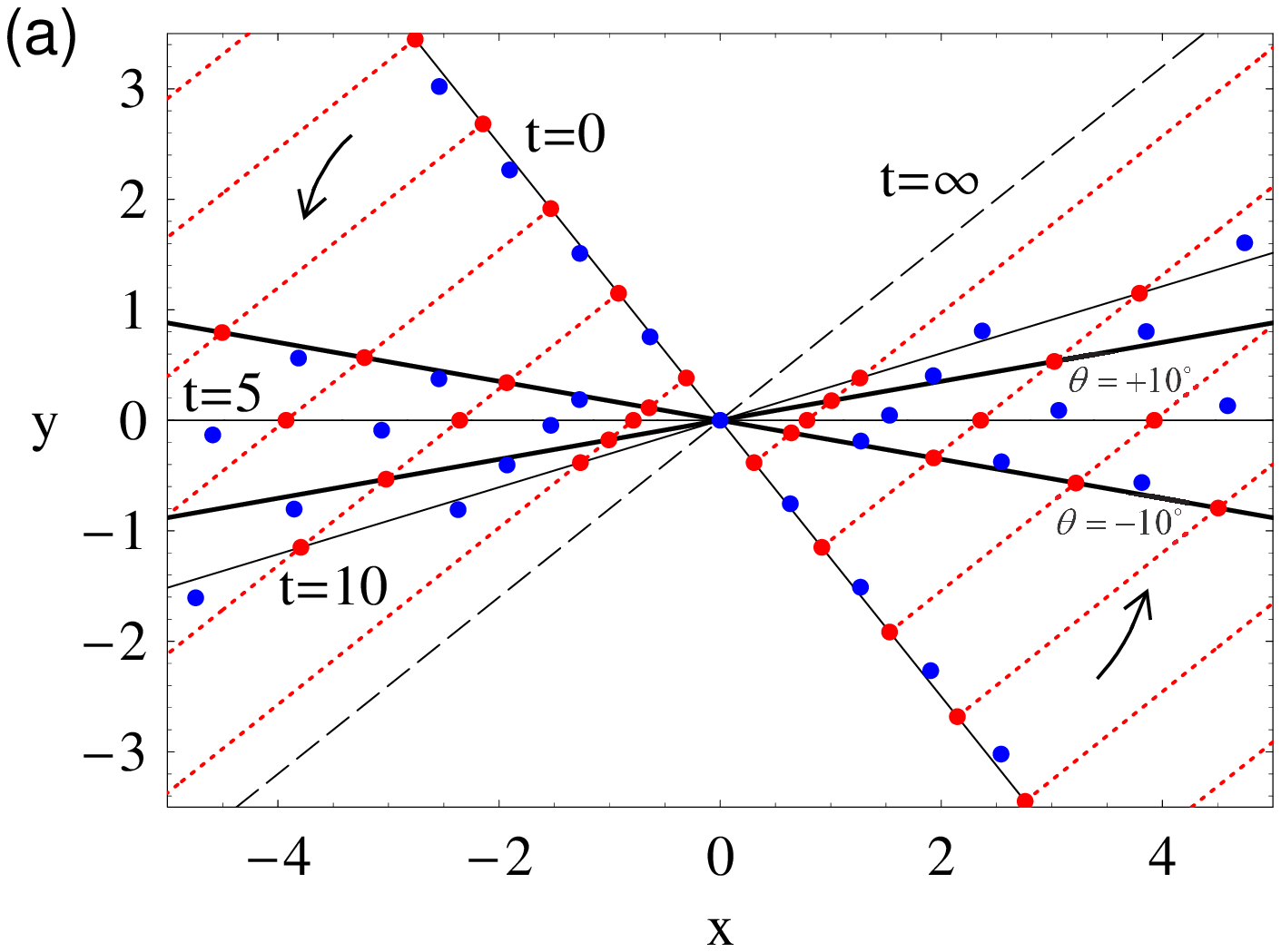}
 \includegraphics[width=6.5cm]{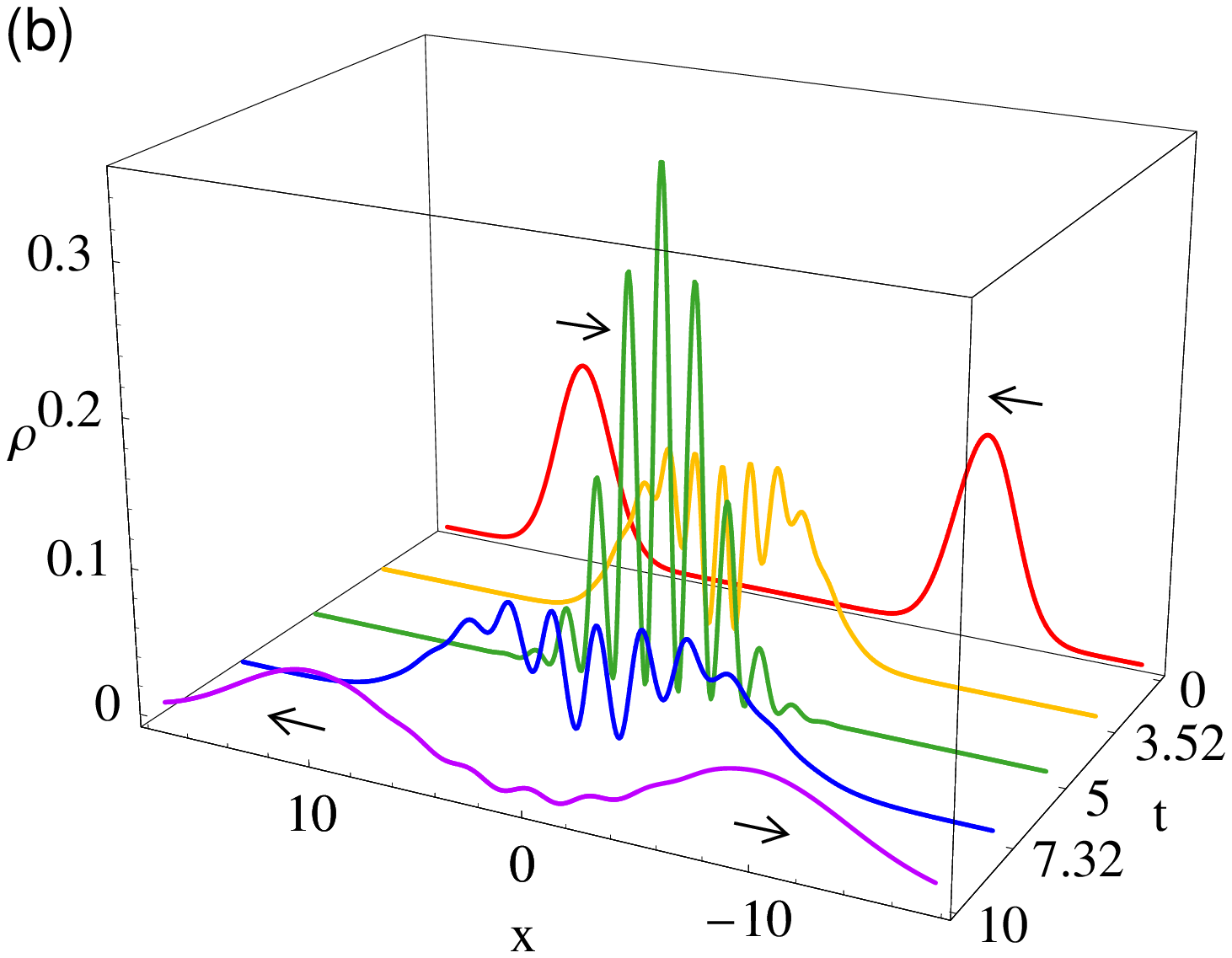}
 \caption{\label{fig8} (Color online)
  (a) Evolution of stagnation points and nodes of the wave function
  with the nodal line (black solid line) and the nodal line at
  $t=\infty$ (black dashed line).
  The arrows indicate the rotational direction of the nodal line.
  Nodal trajectories are shown as dotted lines passing through the
  nodal points.
  Thick black solid lines correspond to the nodal lines with
  $\theta(3.52)=-10^\circ$ and $\theta(7.32)=+10^\circ$.
  Note that $\theta_\infty - \theta_0 = \pi/2$.
  (b) Time-dependence of probability densities along the real axis.}
 \end{center}
\end{figure}


\subsubsection{Rotational dynamics of the nodal line}

As described in Sec.~\ref{sec:Head}, two counter-propagating
Gaussian wave packets interfere with each other at all times in the
complex plane, and the nodal line rotates counterclockwise with
respect to the origin as time progresses. Figure \ref{fig8}(a) shows
the evolution of stagnation points and nodes and nodal trajectories
in the complex plane, and Fig.~\ref{fig8}(b) displays the
time-dependent probability densities along the real axis. Initially,
the interference of tails of two wave packets contributes to the
string of stagnation points and nodes, and the initial nodal line
with the angle $\theta_0=-51.34^\circ$ is perpendicular to nodal
trajectories in Eq.~(\ref{NodalTraj}). Then, the nodal line rotates
counterclockwise and lines up with the real axis at $t=5$. At this
time, the total wave function displays maximal interference and the
exact nodes form on the real axis. After $t=5$, these two wave
packets start to separate but keep interfering with each other in
the complex plane, and the nodal line continues to rotate
counterclockwise away from the real axis. Finally, the angle of the
nodal line tends to the limit angle $\theta_\infty = 38.66^\circ$
when $t$ approaches infinity, and the nodal line becomes parallel to
nodal trajectories. The rotational rate of the nodal line in
Eq.~(\ref{NodalRate}) decays monotonically to zero when $t$ tends to
infinity.  In Fig.~\ref{fig8}(a), the intersections of the nodal
line and nodal trajectories determine the nodal positions, and the
distance between nodes in Eq.~(\ref{NodalDist}) increases with time.
Therefore, the interference process is described by the rotational
dynamics of the nodal line with the angular displacement
$\Delta\theta=\pi/2$.

In conventional quantum mechanics, interference extrema transiently
forming on the real axis are attributed to constructive and
destructive interference between components of the total wave
function, as shown in Fig.~\ref{fig8}(b).  In contrast, in the
complex quantum Hamilton-Jacobi formalism, the interference features
observed on the real axis are described by the counterclockwise
rotation rate of the nodal line in the complex plane.  Since
interference features are observed on the real axis only when the
nodal line is near the real axis, we can define the lifetime for the
interference process corresponding to the time interval for the
nodal line to rotate from $\theta=-10^\circ$ to $\theta=+10^\circ$.
In Fig.~\ref{fig8}(a), the lifetime for the interference features is
$\Delta t=3.8$.  Therefore, compared to the conventional quantum
mechanics, the complex quantum Hamilton-Jacobi formalism provides an
elegant method to define the lifetime for the interference features
observed on the real axis.

\begin{figure}[t]
 \begin{center}
  \includegraphics[width=6.5cm]{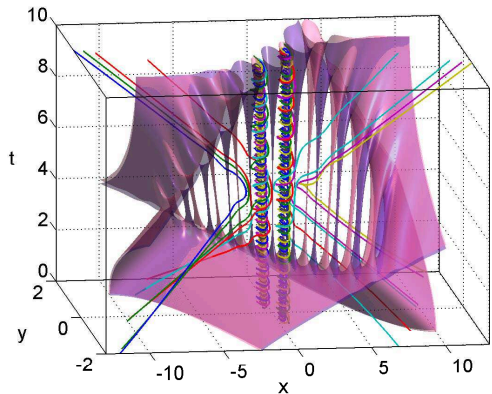}
  \caption{\label{fig9} (Color online)
  P\'olya trajectories displaying helical wrapping around the vortical
  tubes (pink/lighter gray surface) and hyperbolic deflection
  around the stagnation tubes (violet/darker gray sheets).}
 \end{center}
\end{figure}


\subsubsection{Quantum caves with P\'{o}lya trajectories}

Although the QMF displays hyperbolic flow around a node, its associated
PVF displays circular flow \cite{Chou6,Chou7}. Figure~\ref{fig9} shows
that P\'{o}lya trajectories launched from the isochrone arrive at
the real axis with quantum caves. In contrast to Fig.~\ref{fig3}(a),
these trajectories display helical wrapping around the vortical
tubes and hyperbolic deflection around the stagnation tubes.
Both the PVF divergence and vorticity vanish everywhere except at
poles; thus, trajectories display helical wrapping around
\emph{irrotational vortical curves} described by the PVF.


\subsection{Case 2: $v_p \lesssim v_s$}

Next, we consider the case where the relative propagation velocity is
approximately equal to or smaller than the spreading rate of the
wave packets. We use the following initial conditions for Gaussian
wave packets: $x_{0L} = - 5 = - x_{0R}$, $v_{pL} = 1 = -v_{pR}$ and
$\sigma_0 = \sqrt{2}/4$. Maximal interference also occurs at $t=5$
on the real axis and the propagation and spreading velocities are
given by $v_p=1$ and $v_s=\sqrt{2}$, respectively.


\subsubsection{Quantum caves with quantum trajectories and P\'olya
trajectories}

\begin{figure}[t]
 \begin{center}
 \includegraphics[width=6.5cm]{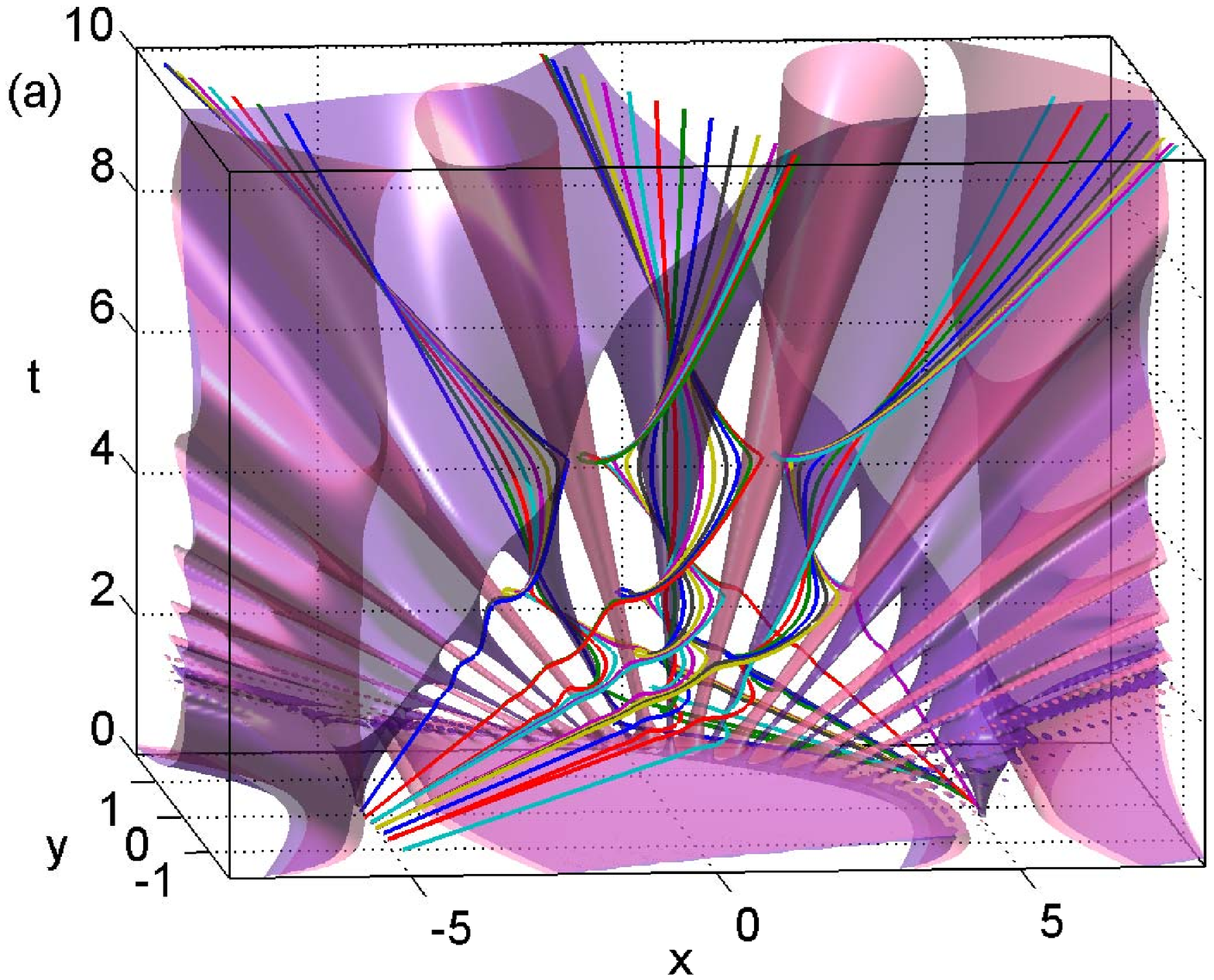}
 \includegraphics[width=6.5cm]{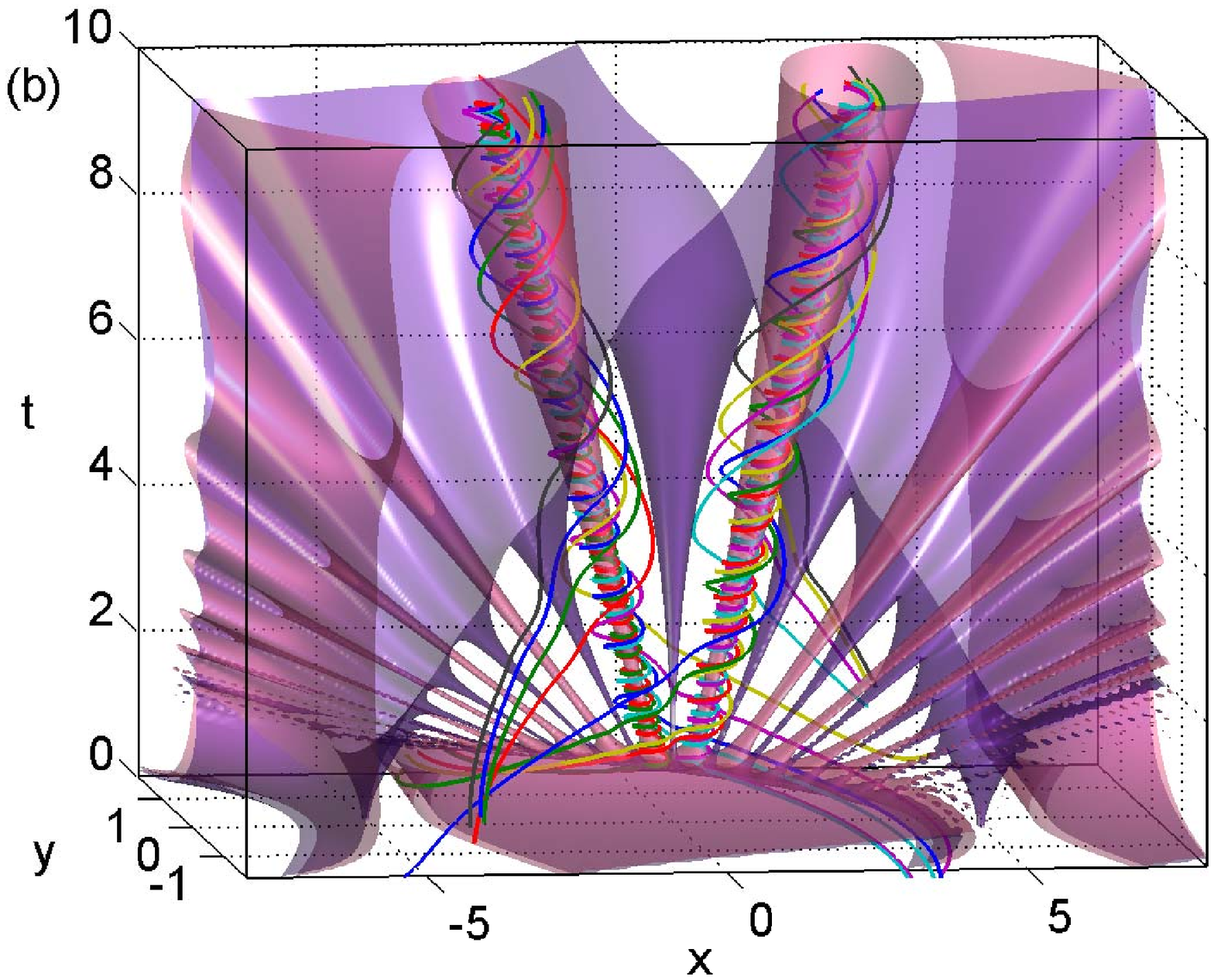}
 \caption{\label{fig10} (Color online)
  Quantum caves (a) with complex quantum trajectories and (b) with
  P\'olya trajectories for the head-on collision of two Gaussian wave
  packets with the relative velocity smaller than the spreading rate.
  These trajectories launched from the isochrone arrive at the real axis at $t=5$.
  The caves are formed with the isosurfaces $|\Psi (z,t)|=0.16$ (pink/lighter gray surface)
  and $|\partial \Psi (z,t)/ \partial z|=0.23$ (violet/darker gray surface).}
 \end{center}
\end{figure}

Figure~\ref{fig10} shows that quantum trajectories and P\'olya
trajectories starting from the isochrone reach the real axis at
$t=5$ with quantum caves consisting of the isosurfaces of the wave
function and its first derivative.  Similar to the case shown in
Fig.~\ref{fig3}, quantum caves form around stagnation curves and
vortical curves appearing alternately, but they are significantly
distorted due to the rapid spreading of the wave packets.  In
addition, quantum trajectories again display helical wrapping around
the stagnation tubes and hyperbolic deflection near the vortical
tubes.  On the contrary, P\'olya trajectories display hyperbolic
deflection near the stagnation tubes and helical wrapping around the
vortical tubes. Again, trajectories launched from different starting
points show long-range correlation, and interference leads to the
formation of quantum caves and produces complicated behavior of
these trajectories.

\subsubsection{Rotational dynamics of the nodal line}

As shown in Fig.~\ref{fig3}, quantum trajectories for Case 1 remain
trapped for a certain time interval between vortical curves, and
then they depart from the stagnation curves.
In contrast, Fig.~\ref{fig10}(a) indicates that quantum trajectories
can wrap around stagnation curves for an infinite time.
This is in agreement with our previous statement that the wrapping
time is a measure of the lifetime of the interference features; in
this case, these features remain visible asymptotically and the
wrapping time becomes infinity.

\begin{figure}[t]
 \begin{center}
 \includegraphics[width=6.5cm]{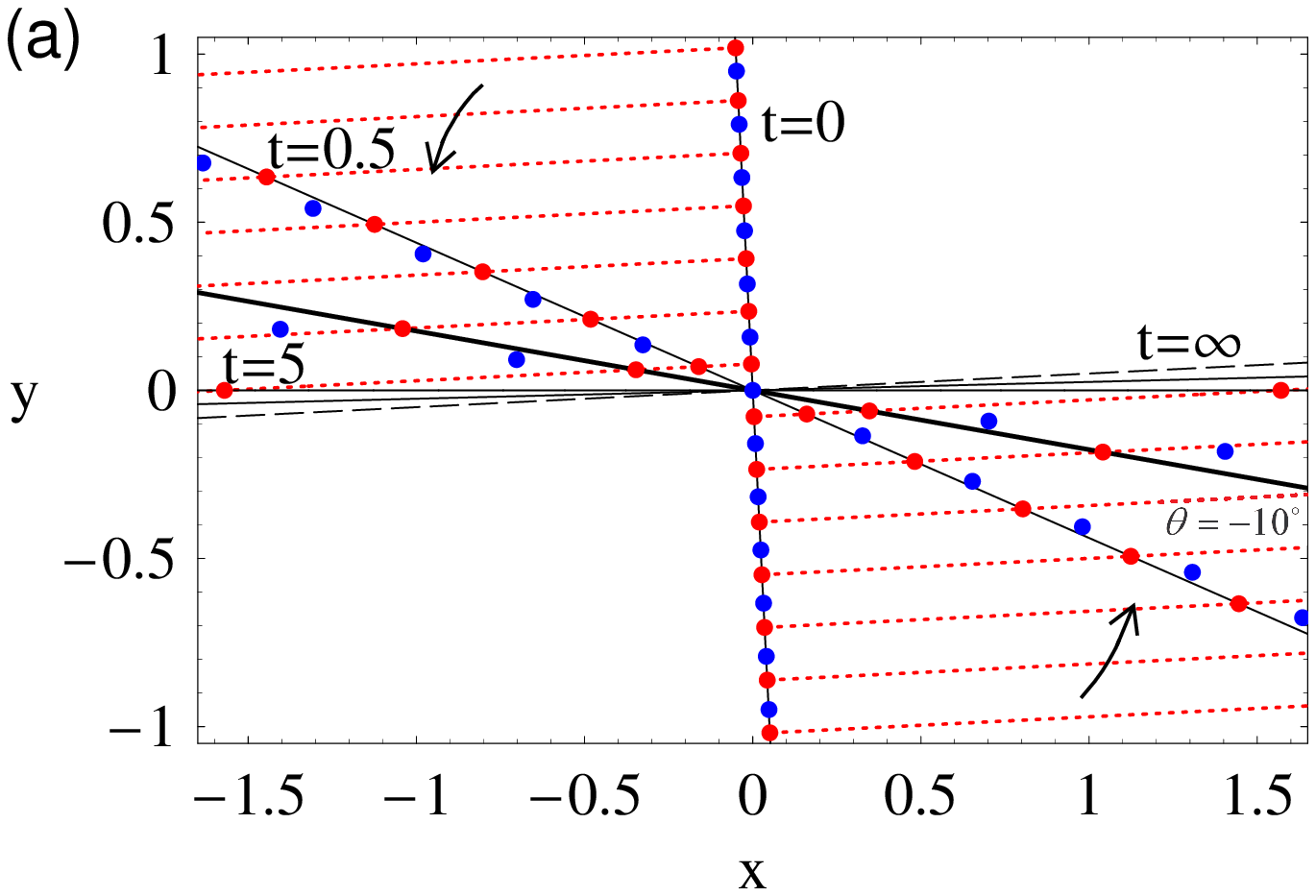}
 \includegraphics[width=6cm]{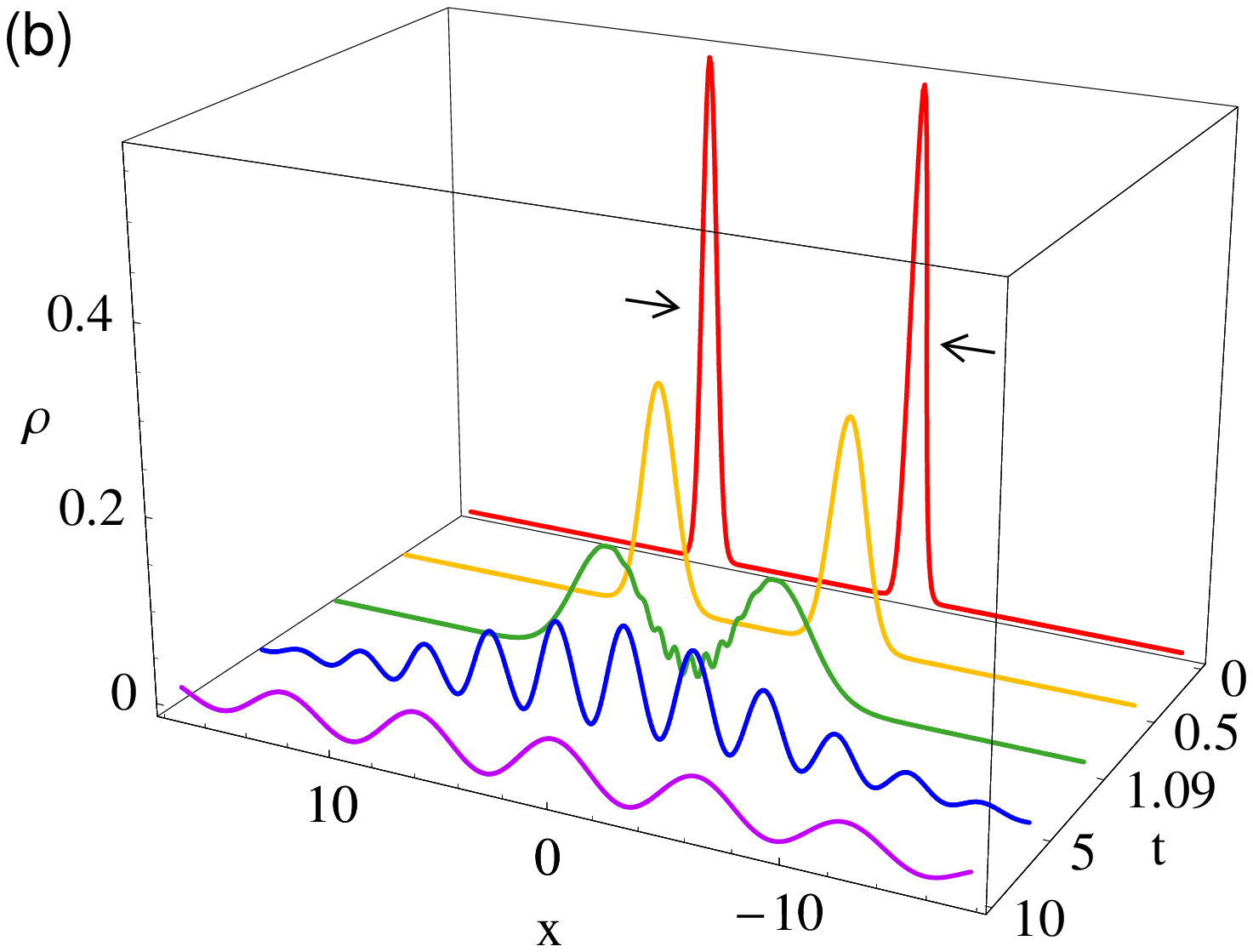}
 \caption{\label{fig11} (Color online)
  (a) Evolution of stagnation points and nodes of the wave function
  with the nodal line (black solid line) and the nodal line at
  $t=\infty$ (black dashed line).
  The arrows indicate the rotational direction of the nodal line.
  Nodal trajectories are shown as dotted lines passing through the
  nodal points.
  Thick black solid lines correspond to the nodal lines with
  $\theta(1.09)=-10^\circ$.
  Note that $\theta_\infty - \theta_0 = \pi/2$.
  (b) Time-dependence of probability densities along the real axis.}
 \end{center}
\end{figure}

Figure~\ref{fig11}(a) shows the evolution of stagnation points and
nodes and nodal trajectories in the complex plane, and
Fig.~\ref{fig11}(b) displays the time-dependent probability
densities along the real axis.  The nodal line starting with the
initial angle $\theta_0 = -87.14^\circ$ rotates counterclockwise
with respect to the origin and reaches the real axis at $t=5$ when
the maximal interference is observed on the real axis.  Then, the
nodal line rotates counterclockwise away from the real axis, and it
approaches the limit nodal line with the angle $\theta_\infty =
2.86^\circ$ as $t$ tends to infinity.  Therefore, the nodal line
rotates with the angular displacement $\Delta\theta=\pi/2$ to reach
the limit nodal line parallel to nodal trajectories.

When the nodal line is near the real axis, the interference features
are clearly displayed on the real axis. As in Case 1, we can define
the starting time of the interference process as the time for the
nodal line with the angle $\theta=-10^\circ$. In
Fig.~\ref{fig11}(a), $\theta(1.09)=-10^\circ$ in this case.  In
addition, the limit nodal line with the angle
$\theta_\infty = 2.86^\circ$ is extremely close to the real axis.
Therefore, the interference process starts at $t=1.09$ and remains
until $t$ tends to infinity. Figure~\ref{fig11}(b) indicates that
the total wave function starts to display the interference feature
when $t=1.09$ and the interference feature persists at long times.

\begin{figure}[t]
 \begin{center}
 \includegraphics[width=6.5cm]{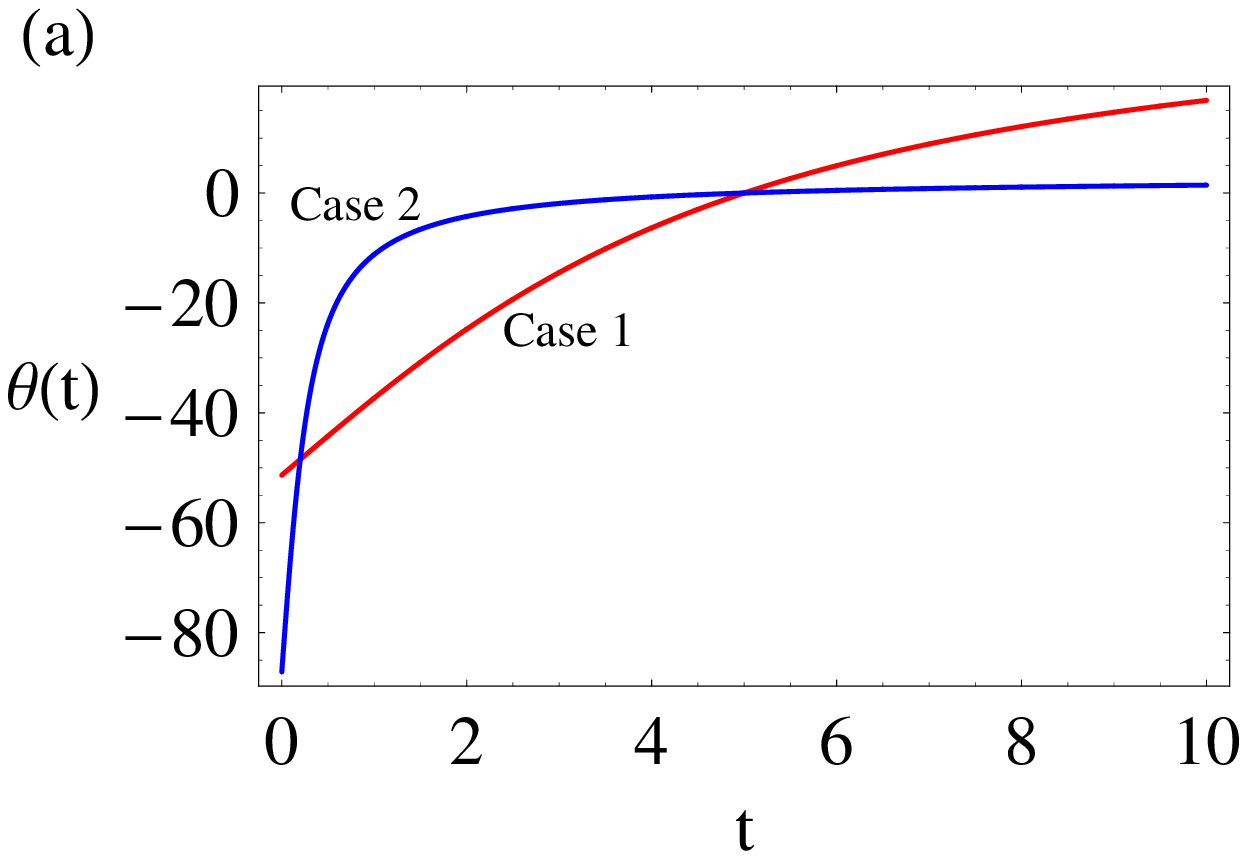}
 \includegraphics[width=6.5cm]{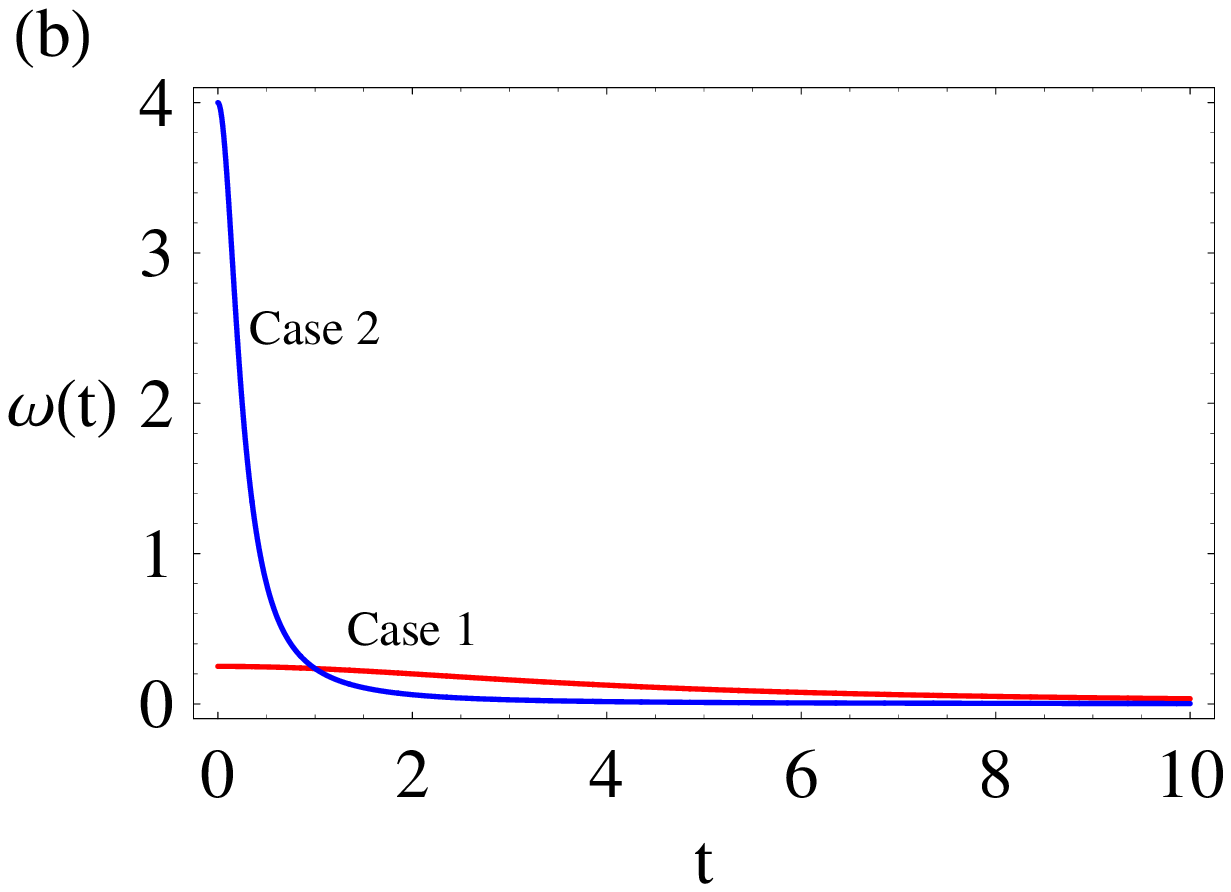}
  \caption{\label{fig12}  (Color online)
  Rotational angle and rotational rate for the nodal line for Case 1
  ($v_p > v_s$) and Case 2 ($v_p \lesssim v_s$):
  (a) Angle of the nodal line in Eq.~(\ref{nodalAngle}).
  (b) Rotational rate of the nodal line in Eq.~(\ref{NodalRate}).}
 \end{center}
\end{figure}

Figure~\ref{fig12} presents the angle and the rotational rate of the
nodal line in Eqs.~(\ref{nodalAngle}) and (\ref{NodalRate}) for Case
1 and Case 2.  As indicated in Eq.~(\ref{NodalRate}), the rotational
rates for these two cases both decay monotonically to zero when $t$
tends to infinity.  For Case 1, the angle of the nodal line alters
relatively slowly and the rotational rate gradually decreases to
zero.  In contrast, for Case 2, the rotational rate shows a rapid
decrease within the initial time interval in Fig.~\ref{fig12}(b).
This dramatic change in the rotational rate reflects the fast
rotation of the nodal line from the initial position to the vicinity
of the real axis in Fig.~\ref{fig12}(a).

\section{\label{sec:Conclusions} Final discussion and concluding
remarks}

In this study, quantum interference was explored in detail within
the complex quantum Hamilton-Jacobi formalism. We reviewed local
structures of the QMF and its associated PVF around stagnation
points and poles, and derived the first-order equation for
approximate quantum trajectories around stagnation points. Analysis
of both the QMF divergence and vorticity along streamlines around a
pole was employed to explain the complicated behavior of complex
quantum trajectories around quantum caves. In addition, both the PVF
divergence and vorticity vanish except at poles; hence, the PVF
describes an incompressible and irrotational flow in the complex
plane. In contrast, both the QMF divergence and vorticity
characterize the turbulent flow in the complex plane.

We analyzed quantum interference using the head-on collision of two
Gaussian wave packets as an example.  Exact detailed analysis was
presented for the rotational dynamics of the nodal line on the real
axis and in the complex plane.  Complex quantum trajectories display
helical wrapping around stagnation tubes and hyperbolic deflection
around vortical tubes.  In contrast, P\'olya trajectories display
hyperbolic deflection around stagnation tubes and helical wrapping
around vortical tubes.  For the case where the relative propagation
velocity is larger than the spreading rate of the wave packets,
during the interference process, trajectories keep circulating
around stagnation tubes as a resonant process and then escape as
time progresses.  Phase singularities in the complex plane can be
regarded as wave dislocations.  Then, the wrapping time for an
individual trajectory was determined by both the QMF divergence and
vorticity, and the average wrapping time was calculated as one of
the definitions for the lifetime of interference.  For the case
where the relative propagation velocity is approximately equal to or
smaller than the spreading rate of the wave packets, the distortion
of quantum caves originates from the rapid spreading of the wave
packets.  Due to the rapid spreading rate, interference features
also develop very rapidly, remaining visible asymptotically in time.
The wrapping time becomes infinity, and this implies the infinite
survival of such interference features.  However, since the
interference features are observed on the real axis only when the
nodal line is near the real axis, the rotational dynamics of the
nodal line in the complex plane offers a unified description to
clearly explain transient or persistent appearance of the
interference features observed on the real axis.  Therefore, these
results show that the complex quantum trajectory method provides a
novel and insightful interpretation of quantum interference.

The average wrapping time determined by both the QMF divergence and
vorticity can be used as one of the definitions for the lifetime of
interference.  On the contrary, the PVF divergence and vorticity
cannot be used to define the lifetime of interference because they
vanish except at poles.  However, the PVF of a complex function,
such as the QMF, contains exactly the same information as the
complex function itself \cite{Needham}.  Thus, it is sufficient to
define the lifetime of interference as the average wrapping time
determined by the QMF divergence and vorticity.

The head-on collision of two Gaussian wave packets with equal
amplitudes was used as a model system to explore quantum
interference in complex space.  This problem is the prototype of
quantum systems displaying interference effects, and it also
exhibits basic features of quantum interference.  A straightforward
generalization is to consider the head-on collision of two Gaussian
wave packets with different amplitudes.  As shown in this paper,
these two wave packets interfere with each other in the complex
plane for all times.  Because of the different amplitudes for these
two counter-propagating wave packets, we cannot observe an infinite
number of nodes on the real axis.  However, nodes can occur on the
real axis at specific times.  In this case, the nodal line displays
not only rotational motion but also translational motion in the
complex plane.  Analogously, the dynamics of the nodal line in the
complex plane clearly explains the interference features observed on
the real axis.  A detailed analysis will be reported in our future
studies.

The current study concentrates mainly on quantum interference
arising from the head-on collision of wave packets involving no
external potential.  We have presented a comprehensive exact
analytical study for this problem, and these results demonstrates
that the complex quantum trajectory method can provide new physical
insights for analyzing, interpreting, and understanding quantum
mechanical problems.  There are various quantum effects resulting
from quantum interference.  Therefore, in the future, quantum
interference incorporating interaction with external potentials
during physical processes can be examined through analytical and
computational approaches.  Multidimensional problems displaying
quantum interference deserve further investigation within the
complex quantum Hamilton-Jacobi formalism.

\section*{Acknowledgment}
Chia-Chun Chou and Robert E.\ Wyatt thank the Robert Welch
Foundation (grant F-0362) for the financial support of this
research. A.\ S.\ Sanz and S.\ Miret-Art\'es acknowledge the
Ministerio de Ciencia e Innovaci\'on (Spain) for financial support
under Project FIS2007-62006. A.\ S.\ Sanz also acknowledges the
Consejo Superior de Investigaciones Cient\'{\i}ficas for a JAE-Doc
contract.


\begin{thebibliography}{99}
\expandafter\ifx\csname natexlab\endcsname\relax\def\natexlab#1{#1}\fi
\expandafter\ifx\csname bibnamefont\endcsname\relax
  \def\bibnamefont#1{#1}\fi
\expandafter\ifx\csname bibfnamefont\endcsname\relax
  \def\bibfnamefont#1{#1}\fi
\expandafter\ifx\csname citenamefont\endcsname\relax
  \def\citenamefont#1{#1}\fi
\expandafter\ifx\csname url\endcsname\relax
  \def\url#1{\texttt{#1}}\fi
\expandafter\ifx\csname urlprefix\endcsname\relax\def\urlprefix{URL }\fi
\providecommand{\bibinfo}[2]{#2}
\providecommand{\eprint}[2][]{\url{#2}}

\bibitem[{\citenamefont{Scalapino}()}]{scalapino}
\bibinfo{author}{\bibfnamefont{D.~J.} \bibnamefont{Scalapino}},
  \bibinfo{note}{in \textit{Tunneling Phenomena in Solids}, edited by E.
  Burstein and S. Lundqvist (Plenum Press, New York, 1969), pp. 477--518.}

\bibitem{friedman}
 J. R. Friedman, V. Patel, W. Chen, S. K. Tolpygo and J. E. Lukens,
 Nature {\bf 406}, 43 (2000).

\bibitem[{\citenamefont{Brumer and Shapiro}(2003)}]{paul}
\bibinfo{author}{\bibfnamefont{P.~W.} \bibnamefont{Brumer}} \bibnamefont{and}
  \bibinfo{author}{\bibfnamefont{M.}~\bibnamefont{Shapiro}},
  \emph{\bibinfo{title}{Principles of the Quantum Control of Molecular
  Processes}} (\bibinfo{publisher}{Wiley-Interscience}, \bibinfo{address}{New
  York}, \bibinfo{year}{2003}).

\bibitem[{ber()}]{berman}
\bibinfo{note}{\textit{Atom Interferometry}, edited by P. R. Berman (Academic
  Press, San Diego, 1997).}

\bibitem{itano}
 D. Leibfried, E. Knill, E. Seidelin, J. Britton, R. B. Blakestad,
 J. Chiaverini, D. B. Hume, W. M. Itano, J. D. Jost, C. Langer,
 R. Ozeri, R. Reichle and D. J. Wineland, Nature {\bf 438}, 639 (2005).

\bibitem{urena1}
 J. O. C\'aceres, M. Morato and A. Gonz\'alez-Ure\~na,
 J. Phys. Chem. A {\bf 110}, 13643 (2006).

\bibitem{urena2}
 A. Gonz\'alez-Ure\~na, A. Requena, A. Bastida and J. Z\'u\~niga,
 Eur. Phys. J. D {\bf 49}, 297 (2008).

\bibitem[{\citenamefont{Chapman et~al.}(1995)\citenamefont{Chapman, Ekstrom,
  Hammond, Schmiedmayer, Tannian, Wehinger, and Pritchard}}]{chapman2}
\bibinfo{author}{\bibfnamefont{M.~S.} \bibnamefont{Chapman}},
  \bibinfo{author}{\bibfnamefont{C.~R.} \bibnamefont{Ekstrom}},
  \bibinfo{author}{\bibfnamefont{T.~D.} \bibnamefont{Hammond}},
  \bibinfo{author}{\bibfnamefont{J.}~\bibnamefont{Schmiedmayer}},
  \bibinfo{author}{\bibfnamefont{B.~E.} \bibnamefont{Tannian}},
  \bibinfo{author}{\bibfnamefont{S.}~\bibnamefont{Wehinger}}, \bibnamefont{and}
  \bibinfo{author}{\bibfnamefont{D.~E.} \bibnamefont{Pritchard}},
  \bibinfo{journal}{Phys. Rev. A} \textbf{\bibinfo{volume}{51}},
  \bibinfo{pages}{R14} (\bibinfo{year}{1995}).

\bibitem[{\citenamefont{Deng et~al.}(1999)\citenamefont{Deng, Hagley,
  Denschlag, Simsarian, Edwards, Clark, Helmerson, Rolston, and
  Phillips}}]{deng}
\bibinfo{author}{\bibfnamefont{L.}~\bibnamefont{Deng}},
  \bibinfo{author}{\bibfnamefont{E.~W.} \bibnamefont{Hagley}},
  \bibinfo{author}{\bibfnamefont{J.}~\bibnamefont{Denschlag}},
  \bibinfo{author}{\bibfnamefont{J.~E.} \bibnamefont{Simsarian}},
  \bibinfo{author}{\bibfnamefont{M.}~\bibnamefont{Edwards}},
  \bibinfo{author}{\bibfnamefont{C.~W.} \bibnamefont{Clark}},
  \bibinfo{author}{\bibfnamefont{K.}~\bibnamefont{Helmerson}},
  \bibinfo{author}{\bibfnamefont{S.~L.} \bibnamefont{Rolston}},
  \bibnamefont{and} \bibinfo{author}{\bibfnamefont{W.~D.}
  \bibnamefont{Phillips}}, \bibinfo{journal}{Phys. Rev. Lett.}
  \textbf{\bibinfo{volume}{83}}, \bibinfo{pages}{5407} (\bibinfo{year}{1999}).

\bibitem{pethick}
 C. J. Pethick and H. Smith, Bose-Einstein Condensation in Dilute
 Gases (Cambridge University Press, Cambridge, 2002).

\bibitem{yoo}
 J. Javanainen and S. M. Yoo, Phys. Rev. Lett. {\bf 76}, 161 (1996).

\bibitem{dalibard}
 Y. Castin and J. Dalibard, Phys. Rev. A {\bf 55}, 4330 (1997).

\bibitem{ketterle}
 M. R. Andrews, C. G. Townsend, H.-J. Miesner, D. S. Durfee,
 D. M. Kurn and W. Ketterle, Science {\bf 275}, 637 (1997).

\bibitem[{\citenamefont{Shin et~al.}(2004)\citenamefont{Shin, Saba, Pasquini,
  Ketterle, Pritchard, and Leanhardt}}]{pritchard}
\bibinfo{author}{\bibfnamefont{Y.}~\bibnamefont{Shin}},
  \bibinfo{author}{\bibfnamefont{M.}~\bibnamefont{Saba}},
  \bibinfo{author}{\bibfnamefont{T.~A.} \bibnamefont{Pasquini}},
  \bibinfo{author}{\bibfnamefont{W.}~\bibnamefont{Ketterle}},
  \bibinfo{author}{\bibfnamefont{D.~E.} \bibnamefont{Pritchard}},
  \bibnamefont{and} \bibinfo{author}{\bibfnamefont{A.~E.}
  \bibnamefont{Leanhardt}}, \bibinfo{journal}{Phys. Rev. Lett.}
  \textbf{\bibinfo{volume}{92}}, \bibinfo{pages}{050405}
  (\bibinfo{year}{2004}).

\bibitem[{\citenamefont{Zhang et~al.}(2006)\citenamefont{Zhang, Zhang, Chapman,
  and You}}]{zhang:070403}
\bibinfo{author}{\bibfnamefont{M.}~\bibnamefont{Zhang}},
  \bibinfo{author}{\bibfnamefont{P.}~\bibnamefont{Zhang}},
  \bibinfo{author}{\bibfnamefont{M.~S.} \bibnamefont{Chapman}},
  \bibnamefont{and} \bibinfo{author}{\bibfnamefont{L.}~\bibnamefont{You}},
  \bibinfo{journal}{Phys. Rev. Lett.} \textbf{\bibinfo{volume}{97}},
  \bibinfo{pages}{070403} (\bibinfo{year}{2006}).

\bibitem[{\citenamefont{Cederbaum et~al.}(2007)\citenamefont{Cederbaum,
  Streltsov, Band, and Alon}}]{cederbaum:110405}
\bibinfo{author}{\bibfnamefont{L.~S.} \bibnamefont{Cederbaum}},
  \bibinfo{author}{\bibfnamefont{A.~I.} \bibnamefont{Streltsov}},
  \bibinfo{author}{\bibfnamefont{Y.~B.} \bibnamefont{Band}}, \bibnamefont{and}
  \bibinfo{author}{\bibfnamefont{O.~E.} \bibnamefont{Alon}},
  \bibinfo{journal}{Phys. Rev. Lett.} \textbf{\bibinfo{volume}{98}},
  \bibinfo{pages}{110405} (\bibinfo{year}{2007}).

\bibitem{chip0}
 T. J. Haigh, A. J. Ferris and M. K. Olsen, arXiv:0907.1333v1 (2009).

\bibitem{chip1}
 T. Schumm, S. Hofferberth, L. M. Andersson, S. Wildermuth, S. Groth,
 I. Bar-Joseph, J. Schmiedmayer and P. Kr\"uger, Nature Physics
 {\bf 1}, 57 (2005).

\bibitem{chip2}
 P. B\"ohl, M. F. Reidel, J. Hoffrogge, J. Reichel, T. W. H\"ansch and
 P. Treutlein, Nature Physics {\bf 5}, 592 (2009).

\bibitem{chip3}
 R. J. Sewell, J. Dingjan, F. Baumg\"artner, I. Llorente-Garc\'{\i}a,
 S. Eriksson, E. A. Hinds, G. Lewis, P. Srinivasan, Z. Moktadir,
 C. O. Gollash and M. Kraft,
 J. Phys. B: At. Mol. Opt. Phys. {\bf 43}, 051003 (2010).

\bibitem{schro}
 J. D. Trimmer, Proc. Am. Phil. Soc. {\bf 124}, 323 (1980).

\bibitem{bohm1}
 D. Bohm, Phys. Rev. {\bf 85}, 166 (1952).

\bibitem{bohm2}
 D. Bohm, Phys. Rev. {\bf 85}, 180 (1952).

\bibitem[{\citenamefont{Holland}(1993)}]{HollandBook}
\bibinfo{author}{\bibfnamefont{P.~R.} \bibnamefont{Holland}},
  \emph{\bibinfo{title}{The Quantum Theory of Motion: An account of the de
  Broglie-Bohm causal interpretation of quantum mechanics}}
  (\bibinfo{publisher}{Cambridge University Press}, \bibinfo{address}{New
  York}, \bibinfo{year}{1993}).

\bibitem[{\citenamefont{Sanz et~al.}(2000)\citenamefont{Sanz, Borondo, and
  Miret-Art\'{e}s}}]{ASSanz}
\bibinfo{author}{\bibfnamefont{A.~S.} \bibnamefont{Sanz}},
  \bibinfo{author}{\bibfnamefont{F.}~\bibnamefont{Borondo}}, \bibnamefont{and}
  \bibinfo{author}{\bibfnamefont{S.}~\bibnamefont{Miret-Art\'{e}s}},
  \bibinfo{journal}{Phys.\ Rev.\ B} \textbf{\bibinfo{volume}{61}},
  \bibinfo{pages}{7743} (\bibinfo{year}{2000}).

\bibitem[{\citenamefont{Sanz et~al.}(2004{\natexlab{a}})\citenamefont{Sanz,
  Borondo, and Miret-Art\'{e}s}}]{ASSanz2}
\bibinfo{author}{\bibfnamefont{A.~S.} \bibnamefont{Sanz}},
  \bibinfo{author}{\bibfnamefont{F.}~\bibnamefont{Borondo}}, \bibnamefont{and}
  \bibinfo{author}{\bibfnamefont{S.}~\bibnamefont{Miret-Art\'{e}s}},
  \bibinfo{journal}{Phys.\ Rev.\ B} \textbf{\bibinfo{volume}{69}},
  \bibinfo{pages}{115413} (\bibinfo{year}{2004}{\natexlab{a}}).

\bibitem[{\citenamefont{Guantes et~al.}(2004)\citenamefont{Guantes, Sanz,
  Margalef-Roig, and Miret-Art\'{e}s}}]{ASSanz4}
\bibinfo{author}{\bibfnamefont{R.}~\bibnamefont{Guantes}},
  \bibinfo{author}{\bibfnamefont{A.~S.} \bibnamefont{Sanz}},
  \bibinfo{author}{\bibfnamefont{J.}~\bibnamefont{Margalef-Roig}},
  \bibnamefont{and}
  \bibinfo{author}{\bibfnamefont{S.}~\bibnamefont{Miret-Art\'{e}s}},
  \bibinfo{journal}{Surf.\ Sci.\ Rep.} \textbf{\bibinfo{volume}{53}},
  \bibinfo{pages}{199} (\bibinfo{year}{2004}).

\bibitem[{\citenamefont{Sanz and
  Miret-Art\'{e}s}(2007{\natexlab{a}})}]{ASSanz6}
\bibinfo{author}{\bibfnamefont{A.~S.} \bibnamefont{Sanz}} \bibnamefont{and}
  \bibinfo{author}{\bibfnamefont{S.}~\bibnamefont{Miret-Art\'{e}s}},
  \bibinfo{journal}{J.\ Chem.\ Phys.} \textbf{\bibinfo{volume}{126}},
  \bibinfo{pages}{234106} (\bibinfo{year}{2007}{\natexlab{a}}).

\bibitem[{\citenamefont{Sanz and
  Miret-Art\'{e}s}(2007{\natexlab{b}})}]{ASSanz7}
\bibinfo{author}{\bibfnamefont{A.~S.} \bibnamefont{Sanz}} \bibnamefont{and}
  \bibinfo{author}{\bibfnamefont{S.}~\bibnamefont{Miret-Art\'{e}s}},
  \bibinfo{journal}{Chem.\ Phys.\ Lett.} \textbf{\bibinfo{volume}{445}},
  \bibinfo{pages}{350} (\bibinfo{year}{2007}{\natexlab{b}}).

\bibitem[{\citenamefont{Sanz and
  Miret-Art\'{e}s}(2008{\natexlab{a}})}]{ASSanz9}
\bibinfo{author}{\bibfnamefont{A.~S.} \bibnamefont{Sanz}} \bibnamefont{and}
  \bibinfo{author}{\bibfnamefont{S.}~\bibnamefont{Miret-Art\'{e}s}},
  \bibinfo{journal}{J.\ Phys.\ A: Math.\ Theor.} \textbf{\bibinfo{volume}{41}},
  \bibinfo{pages}{435303} (\bibinfo{year}{2008}{\natexlab{a}}).

\bibitem[{\citenamefont{Lopreore and Wyatt}(1999)}]{LopreoreWyatt1}
\bibinfo{author}{\bibfnamefont{C.~L.} \bibnamefont{Lopreore}} \bibnamefont{and}
  \bibinfo{author}{\bibfnamefont{R.~E.} \bibnamefont{Wyatt}},
  \bibinfo{journal}{Phys.\ Rev.\ Lett.} \textbf{\bibinfo{volume}{82}},
  \bibinfo{pages}{5190} (\bibinfo{year}{1999}).

\bibitem[{\citenamefont{Wyatt}(2005)}]{WyattBook}
\bibinfo{author}{\bibfnamefont{R.~E.} \bibnamefont{Wyatt}},
  \emph{\bibinfo{title}{Quantum Dynamics with Trajectories: Introduction to
  quantum hydrodynamics}} (\bibinfo{publisher}{Springer}, \bibinfo{address}{New
  York}, \bibinfo{year}{2005}).

\bibitem[{Poirier(2004)}]{Poirier1}
\bibinfo{author}{B.~Poirier}, \bibinfo{journal}{J.\ Chem.\ Phys.}
  \textbf{\bibinfo{volume}{121}}, \bibinfo{pages}{4501}
  (\bibinfo{year}{2004}).

\bibitem[{Trahan and Poirier(2006{\natexlab{a}})}]{Poirier2}
\bibinfo{author}{C.~Trahan}, \bibinfo{author}{B.~Poirier},
  \bibinfo{journal}{J.\ Chem.\ Phys.} \textbf{\bibinfo{volume}{124}},
  \bibinfo{pages}{034115} (\bibinfo{year}{2006}{\natexlab{a}}).

\bibitem[{Trahan and Poirier(2006{\natexlab{b}})}]{Poirier3}
\bibinfo{author}{C.~Trahan}, \bibinfo{author}{B.~Poirier},
  \bibinfo{journal}{J.\ Chem.\ Phys.} \textbf{\bibinfo{volume}{124}},
  \bibinfo{pages}{034116} (\bibinfo{year}{2006}{\natexlab{b}}).

\bibitem[{Poirier and Parlant(2007)}]{Poirier5}
\bibinfo{author}{B.~Poirier}, \bibinfo{author}{G.~Parlant},
  \bibinfo{journal}{J.\ Phys.\ Chem.\ A} \textbf{\bibinfo{volume}{111}},
  \bibinfo{pages}{10400} (\bibinfo{year}{2007}).

\bibitem[{Poirier(2008{\natexlab{a}})}]{Poirier6}
\bibinfo{author}{B.~Poirier}, \bibinfo{journal}{J.\ Chem.\ Phys.}
  \textbf{\bibinfo{volume}{128}}, \bibinfo{pages}{164115}
  (\bibinfo{year}{2008}{\natexlab{a}}).

\bibitem[{Poirier(2008{\natexlab{b}})}]{Poirier7}
\bibinfo{author}{B.~Poirier}, \bibinfo{journal}{J.\ Chem.\ Phys.}
  \textbf{\bibinfo{volume}{129}}, \bibinfo{pages}{084103}
 (\bibinfo{year}{2008}{\natexlab{b}}).

\bibitem[{Park et~al.(2008)Park, Poirier, and Parlant}]{Poirier8}
\bibinfo{author}{K.~Park}, \bibinfo{author}{B.~Poirier},
  \bibinfo{author}{G.~Parlant}, \bibinfo{journal}{J.\ Chem.\ Phys.}
  \textbf{\bibinfo{volume}{129}}, \bibinfo{pages}{194112}
  (\bibinfo{year}{2008}).

\bibitem{LP1}
 R. A. Leacock and M. J. Padgett, Phys. Rev. Lett. {\bf 50}, 3 (1983).

\bibitem{LP2}
 R. A. Leacock and M. J. Padgett, Phys. Rev. D {\bf 28}, 2491 (1983).

\bibitem{MJohn1}
 M. V. John, Found. Phys. Lett. {\bf 15}, 329 (2002).

\bibitem{MJohn2}
 M. V. John, Ann. Phys. {\bf 324}, 220 (2009).

\bibitem{CDYang1}
 C. D. Yang, Ann. Phys. (N.Y.) {\bf 319}, 444 (2005).

\bibitem{CDYang2}
 C. D. Yang, Ann. Phys. (N.Y.) {\bf 321}, 2876 (2006).

\bibitem{CDYang3}
 C. D. Yang, Chaos Soliton Fract. {\bf 30}, 342 (2006).

\bibitem{Chou3a}
 C.-C. Chou and R. E. Wyatt, Phys. Rev. A {\bf 76}, 012115 (2007).

\bibitem{Chou3b}
 C.-C. Chou and R. E. Wyatt, J. Chem. Phys. {\bf 128}, 154106 (2008).

\bibitem{Tannor2a}
 Y. Goldfarb, I. Degani and D. J. Tannor,
 J. Chem. Phys. {\bf 125}, 231103 (2006).

\bibitem{Tannor2b}
 A. S. Sanz and S. Miret-Art\'{e}s, J. Chem. Phys. {\bf 127}, 197101 (2007).

\bibitem{Tannor2c}
 Y. Goldfarb, I. Degani and D. J. Tannor,
 J. Chem. Phys. {\bf 127}, 197102 (2007).

\bibitem{Tannor3a}
 Y. Goldfarb, J. Schiff and D. J. Tannor,
 J. Phys. Chem. A {\bf 111}, 10416 (2007).

\bibitem{Tannor3b}
 Y. Goldfarb and D. J. Tannor, J. Chem. Phys. {\bf 127}, 161101 (2007).

\bibitem{Rowland1a}
 B. A. Rowland and R. E. Wyatt,
 J. Phys. Chem. A {\bf 111}, 10234 (2007).

\bibitem{Rowland1b}
 B. A. Rowland and R. E. Wyatt,
 J. Chem. Phys. {\bf 127}, 164104 (2007).

\bibitem{Rowland1c}
 B. A. Rowland and R. E. Wyatt,
 Chem. Phys. Lett. {\bf 461}, 155 (2008).

\bibitem{Rowland2a}
 R. E. Wyatt and B. A. Rowland,
 J. Chem. Phys. {\bf 127}, 044103 (2007);.

\bibitem{Rowland2b}
 R. E. Wyatt and B. A. Rowland,
 J. Chem. Theory Comput. {\bf 5}, 443 (2009).

\bibitem{Rowland2c}
 R. E. Wyatt and B. A. Rowland,
 J. Chem. Theory Comput. {\bf 5}, 452 (2009).

\bibitem[{\citenamefont{Sanz and
  Miret-Art\'{e}s}(2008{\natexlab{b}})}]{ASSanz8}
\bibinfo{author}{\bibfnamefont{A.~S.} \bibnamefont{Sanz}} \bibnamefont{and}
  \bibinfo{author}{\bibfnamefont{S.}~\bibnamefont{Miret-Art\'{e}s}},
  \bibinfo{journal}{Chem.\ Phys.\ Lett.} \textbf{\bibinfo{volume}{458}},
  \bibinfo{pages}{239} (\bibinfo{year}{2008}{\natexlab{b}}).

\bibitem{haensel-a}
 W. H\"ansel, J. Reichel, P. Hommelhoff and T. W. H\"ansch,
 Phys. Rev. Lett. {\bf 86}, 608 (2001).

\bibitem{haensel-b}
 W. H\"ansel, J. Reichel, P. Hommelhoff and T. W. H\"ansch,
 Phys. Rev. A {\bf 64}, 063607 (2001).

\bibitem{haensel-c}
 E. A. Hinds, C. J. Vale and M. G. Boshier,
 Phys. Rev. Lett. {\bf 86}, 1462 (2001).

\bibitem{haensel-d}
 E. Andersson, T. Calarco, R. Folman, M. Andersson, B. Hessmo and
 J. Schmiedmayer, Phys. Rev. Lett. {\bf 88}, 100401 (2002).

\bibitem{haensel-e}
 K. T. Kapale and J. P. Dowling,
 Phys. Rev. Lett. {\bf 95}, 173601 (2005).

\bibitem[{\citenamefont{Chou et~al.}(2009)\citenamefont{Chou, Sanz,
  Miret-Art\'{e}s, and Wyatt}}]{Chou9}
\bibinfo{author}{\bibfnamefont{C.-C.} \bibnamefont{Chou}},
  \bibinfo{author}{\bibfnamefont{A.~S.} \bibnamefont{Sanz}},
  \bibinfo{author}{\bibfnamefont{S.}~\bibnamefont{Miret-Art\'{e}s}},
  \bibnamefont{and} \bibinfo{author}{\bibfnamefont{R.~E.} \bibnamefont{Wyatt}},
  \bibinfo{journal}{Phys.~Rev.~Lett.} \textbf{\bibinfo{volume}{102}},
  \bibinfo{pages}{250401} (\bibinfo{year}{2009}).

\bibitem[{\citenamefont{Chou and Wyatt}(2008{\natexlab{b}})}]{Chou6}
\bibinfo{author}{\bibfnamefont{C.-C.} \bibnamefont{Chou}} \bibnamefont{and}
  \bibinfo{author}{\bibfnamefont{R.~E.} \bibnamefont{Wyatt}},
  \bibinfo{journal}{J.\ Chem.\ Phys.} \textbf{\bibinfo{volume}{128}},
  \bibinfo{pages}{234106} (\bibinfo{year}{2008}{\natexlab{b}}).

\bibitem[{\citenamefont{Chou and Wyatt}(2008{\natexlab{c}})}]{Chou7}
\bibinfo{author}{\bibfnamefont{C.-C.} \bibnamefont{Chou}} \bibnamefont{and}
  \bibinfo{author}{\bibfnamefont{R.~E.} \bibnamefont{Wyatt}},
  \bibinfo{journal}{J.\ Chem.\ Phys.} \textbf{\bibinfo{volume}{129}},
  \bibinfo{pages}{124113} (\bibinfo{year}{2008}{\natexlab{c}}).

\bibitem[{\citenamefont{Needham}(1998)}]{Needham}
\bibinfo{author}{\bibfnamefont{T.}~\bibnamefont{Needham}},
  \emph{\bibinfo{title}{Visual Complex Analysis}} (\bibinfo{publisher}{Oxford
  University Press}, \bibinfo{address}{Oxford}, \bibinfo{year}{1998}).

\bibitem[{\citenamefont{P\'{o}lya and Latta}(1974)}]{Polya}
\bibinfo{author}{\bibfnamefont{G.}~\bibnamefont{P\'{o}lya}} \bibnamefont{and}
  \bibinfo{author}{\bibfnamefont{G.}~\bibnamefont{Latta}},
  \emph{\bibinfo{title}{Complex Variables}} (\bibinfo{publisher}{John Wiley \&
  Sons}, \bibinfo{address}{New York}, \bibinfo{year}{1974}).

\bibitem[{\citenamefont{Braden}(1987)}]{Braden}
\bibinfo{author}{\bibfnamefont{B.}~\bibnamefont{Braden}},
  \bibinfo{journal}{Math.\ Mag.} \textbf{\bibinfo{volume}{60}},
  \bibinfo{pages}{321} (\bibinfo{year}{1987}).

\bibitem[{\citenamefont{Dirac}(1931)}]{Dirac}
\bibinfo{author}{\bibfnamefont{P.~A.} \bibnamefont{Dirac}},
  \bibinfo{journal}{Proc.\ R.\ Soc.\ A} \textbf{\bibinfo{volume}{133}},
  \bibinfo{pages}{60} (\bibinfo{year}{1931}).

\bibitem{McCullough1}
 E. A. McCullough, Jr. and R. E. Wyatt,
 J. Chem. Phys. {\bf 51}, 1253 (1969).

\bibitem{McCullough2}
 E. A. McCullough, Jr. and R. E. Wyatt,
 J. Chem. Phys. {\bf 54}, 3578 (1971).

\bibitem{McCullough3}
 E. A. McCullough, Jr. and R. E. Wyatt,
 J. Chem. Phys. {\bf 54}, 3592 (1971).

\bibitem{Hirschfelder1}
 J. O. Hirschfelder, A. C. Christoph and W. E. Palke,
 J. Chem. Phys. {\bf 61}, 5435 (1974).

\bibitem{Hirschfelder2}
 J. O. Hirschfelder, C. J. Goebel and L. W. Bruch,
 J. Chem. Phys. {\bf 61}, 5456 (1974).

\bibitem{Hirschfelder3}
 J. O. Hirschfelder and K. T. Tang,
 J. Chem. Phys. {\bf 64}, 760 (1976).

\bibitem{Hirschfelder4}
 J. O. Hirschfelder and K. T. Tang,
 J. Chem. Phys. {\bf 65}, 470 (1976).

\bibitem{Hirschfelder5}
 J. O. Hirschfelder, J. Chem. Phys. {\bf 67}, 5477 (1977).

\bibitem[{\citenamefont{Sanz et~al.}(2004{\natexlab{b}})\citenamefont{Sanz,
  Borondo, and Miret-Art\'{e}s}}]{ASSanz3}
\bibinfo{author}{\bibfnamefont{A.~S.} \bibnamefont{Sanz}},
  \bibinfo{author}{\bibfnamefont{F.}~\bibnamefont{Borondo}}, \bibnamefont{and}
  \bibinfo{author}{\bibfnamefont{S.}~\bibnamefont{Miret-Art\'{e}s}},
  \bibinfo{journal}{J.\ Chem.\ Phys.} \textbf{\bibinfo{volume}{120}},
  \bibinfo{pages}{8794} (\bibinfo{year}{2004}{\natexlab{b}}).

\bibitem{Chou8a}
 C.-C. Chou and R. E. Wyatt, Phys. Rev. A {\bf 78}, 044101 (2008).

\bibitem{Chou8b}
 C.-C. Chou and R. E. Wyatt, Phys. Lett. A {\bf 373}, 1811 (2009).

\bibitem[{\citenamefont{Hirsch et~al.}(2003)\citenamefont{Hirsch, Smale, and
  Devaney}}]{Hirsch}
\bibinfo{author}{\bibfnamefont{M.~W.} \bibnamefont{Hirsch}},
  \bibinfo{author}{\bibfnamefont{S.}~\bibnamefont{Smale}}, \bibnamefont{and}
  \bibinfo{author}{\bibfnamefont{R.~L.} \bibnamefont{Devaney}},
  \emph{\bibinfo{title}{Differential Equations, Dynamical Systems, and an
  Introduction to Chaos}} (\bibinfo{publisher}{Academic Press},
  \bibinfo{address}{San Diego}, \bibinfo{year}{2003}).

\bibitem[{\citenamefont{Tritton}(1988)}]{Tritton}
\bibinfo{author}{\bibfnamefont{D.~J.} \bibnamefont{Tritton}},
  \emph{\bibinfo{title}{Physical Fluid Dynamics}} (\bibinfo{publisher}{Oxford
  University Press}, \bibinfo{address}{Oxford}, \bibinfo{year}{1988}).

\bibitem[{\citenamefont{Nye and Berry}(1974)}]{NyeBerry}
\bibinfo{author}{\bibfnamefont{J.~F.} \bibnamefont{Nye}} \bibnamefont{and}
  \bibinfo{author}{\bibfnamefont{M.~V.} \bibnamefont{Berry}},
  \bibinfo{journal}{Proc.\ R.\ Soc. A} \textbf{\bibinfo{volume}{336}},
  \bibinfo{pages}{165} (\bibinfo{year}{1974}).

\end{thebibliography}

\end{document}